\documentclass[10pt,sigconf,nonacm=true]{acmart}
\usepackage{paralist}
\usepackage{xspace}
\usepackage{comment}
\usepackage{epsfig}
\PassOptionsToPackage{hyphens}{url}
\usepackage{color}

\usepackage{ragged2e}
\usepackage{ifthen}
\usepackage{amsmath}
\usepackage{subfig}
\usepackage{tikz}
\usepackage{multicol}
\usepackage{xcolor}
\usepackage{multirow}
\newcommand{\sys}{MaxMem\xspace}
\newcommand{\libsys}{libMaxMem\xspace}
\newcommand{\papertitle}{\huge \sys: Colocation and Performance for
  Big Data Applications on Tiered Main Memory Servers}

\usepackage[normalem]{ulem}

\usepackage{enumitem}

\newcommand*\circled[1]{\tikz[baseline=(char.base)]{
            \node[shape=circle,fill,inner sep=.2pt] (char)
            {\textcolor{white}{\small \\#1}};}}

\begin{document}

% \clubpenalty=10000
% \widowpenalty = 10000

% \renewcommand\footnotetextcopyrightpermission[1]{}
% \pagestyle{plain}

% \ifpdf
% \setlength{\pdfpagewidth}{8.5in}
% \setlength{\pdfpageheight}{11in}
% \else
% \fi

% \setlength{\belowcaptionskip}{-10pt}
% \setlength{\textfloatsep}{12.0pt}

% \title{\Large \bf \sys: Efficient Tiered Memory Management is in the
%   Operating System}
% \title{\Large \bf \sys: Efficient Tiered Main Memory Management}
% \title{\Large \bf \sys: Software Tiered Main Memory Management for
%   Big Data Applications}
% \title{\large \sys: Software Tiered Main Memory Management for Big
%   Data Applications\vspace{-1.5cm}}
\title{\papertitle}
% \author{Paper \#216, \pageref{LastPage} pages}
% \author{Paper \#xxx}
% \author{Anonymous submission \#90}

%\author{\normalsize Anonymous submission \#666}

\author{Amanda Raybuck}
\affiliation{\institution{The University of Texas at Austin}\city{}\country{}}

\author{Wei Zhang}
\affiliation{\institution{Microsoft}\city{}\country{}}

\author{Kayvan Mansoorshahi}
\affiliation{\institution{The University of Texas at Austin}\city{}\country{}}

\author{Aditya K Kamath}
\affiliation{\institution{University of Washington}\city{}\country{}}

\author{Mattan Erez}
\affiliation{\institution{The University of Texas at Austin}\city{}\country{}}

\author{Simon Peter}
\affiliation{\institution{University of Washington}\city{}\country{}}

\date{}

\begin{abstract}
	% Tiered memory systems need to be QoS-aware to provide balanced
% performance on consolidated servers, while handling dynamically
% changing application demands. However, existing tiered memory systems
% only provide coarse-grained or static QoS among applications. We
% explore the performance implications of application colocation with
% current systems and show that application tail latency and throughput
% are affected by up to 35\% and 12\% versus isolation, respectively.

We present \sys, a tiered main memory management system that aims to
maximize Big Data application colocation and performance. \sys uses an
application-agnostic and lightweight memory occupancy control
mechanism based on fast memory miss ratios to provide application QoS
under increasing colocation. By relying on memory access sampling and
binning to quickly identify per-process memory heat gradients, \sys
maximizes performance for many applications sharing tiered main memory
simultaneously. \sys is designed as a user-space memory manager to be
easily modifiable and extensible, without complex kernel code
development. On a system with tiered main memory consisting of DRAM
and Intel Optane persistent memory modules, our evaluation confirms
that \sys provides 11\% and 38\% better throughput and up to 80\% and
an order of magnitude lower 99th percentile latency than HeMem and 
Linux AutoNUMA, respectively, with a Big Data key-value store in 
dynamic colocation scenarios.

\end{abstract}

% \settopmatter{printacmref=false, printfolios=true}
\maketitle

\newcommand{\tts}[1]{\texttt{\small #1}}

\definecolor{Orange}{rgb}{1,0.5,0}
\definecolor{Gray}{rgb}{0.5,0.5,0.5}
\definecolor{LtGray}{rgb}{0.66,0.66,0.66}
\definecolor{Red}{rgb}{0.8,0,0}
\definecolor{Green}{rgb}{0,0.33,0}

\newcommand{\brok}[1]{\sout{#1}}
\newcommand{\cbrok}[1]{\textcolor{Gray}{\brok{#1}}}
\newcommand{\less}[1]{\ensuremath{\check{#1}}}
\newcommand{\cless}[1]{\textcolor{Gray}{\less{#1}}}
\newcommand{\ok}[1]{#1}
\newcommand{\cok}[1]{\ok{#1}}

\renewcommand{\b}[1]{\cbrok{#1}}
\renewcommand{\l}[1]{\cless{#1}}
\renewcommand{\o}[1]{\cok{#1}}

\newcommand{\vhead}[1]{\begin{sideways}#1\end{sideways}}
\newcommand{\none}{-}
\newcommand{\broken}{\ensuremath{\otimes}}
\newcommand{\lessfunc}{\ensuremath{\circleddash}}
\newcommand{\phone}{Phone \#}
\newcommand{\histbookmarks}{Bookmarks}
\newcommand{\sms}{SMS}
\newcommand{\imei}{Device ID}
\newcommand{\location}{Location}
\newcommand{\contacts}{Contacts}
\newcommand{\calendar}{Calendar}
\newcommand{\verify}[1]{\textcolor{Red}{#1}}

\newcommand{\sur}[0]{S}
\newcommand{\cov}[0]{C}
\newcommand{\ove}[0]{O}

\newcommand{\ooo}[0]{\o{\sur}\o{\cov}\o{\ove}}
\newcommand{\ool}[0]{\o{\sur}\o{\cov}\l{\ove}}
\newcommand{\oob}[0]{\o{\sur}\o{\cov}\b{\ove}}
\newcommand{\olo}[0]{\o{\sur}\l{\cov}\o{\ove}}
\newcommand{\oll}[0]{\o{\sur}\l{\cov}\l{\ove}}
\newcommand{\olb}[0]{\o{\sur}\l{\cov}\b{\ove}}
\newcommand{\obo}[0]{\o{\sur}\b{\cov}\o{\ove}}
\newcommand{\obl}[0]{\o{\sur}\b{\cov}\l{\ove}}
\newcommand{\obb}[0]{\o{\sur}\b{\cov}\b{\ove}}

\newcommand{\loo}[0]{\l{\sur}\o{\cov}\o{\ove}}
\newcommand{\lol}[0]{\l{\sur}\o{\cov}\l{\ove}}
\newcommand{\lob}[0]{\l{\sur}\o{\cov}\b{\ove}}
\newcommand{\llo}[0]{\l{\sur}\l{\cov}\o{\ove}}
\newcommand{\llll}[0]{\l{\sur}\l{\cov}\l{\ove}} % orange?
\newcommand{\llb}[0]{\l{\sur}\l{\cov}\b{\ove}} % orange?
\newcommand{\lbo}[0]{\l{\sur}\b{\cov}\o{\ove}}
\newcommand{\lbl}[0]{\l{\sur}\b{\cov}\l{\ove}} % orange?
\newcommand{\lbb}[0]{\l{\sur}\b{\cov}\b{\ove}} % orange?

\newcommand{\boo}[0]{\b{\sur}\o{\cov}\o{\ove}}
\newcommand{\bol}[0]{\b{\sur}\o{\cov}\l{\ove}}
\newcommand{\bob}[0]{\b{\sur}\o{\cov}\b{\ove}}
\newcommand{\blo}[0]{\b{\sur}\l{\cov}\o{\ove}}
\newcommand{\bll}[0]{\b{\sur}\l{\cov}\l{\ove}}  % orange?
\newcommand{\blb}[0]{\b{\sur}\l{\cov}\b{\ove}}  % orange?
\newcommand{\bbo}[0]{\b{\sur}\b{\cov}\o{\ove}}
\newcommand{\bbl}[0]{\b{\sur}\b{\cov}\l{\ove}}  % orange?
\newcommand{\bbb}[0]{\textcolor{red}{\brok{\sur}\brok{\cov}\brok{\ove}}}

\newcommand{\vladd}[1]{\textcolor{cyan}{#1}}
\newcommand{\vlremove}[1]{\textcolor{Gray}{\sout{#1}}}
\newcommand{\vlreplace}[2]{\textcolor{Gray}{\sout{#1}}\textcolor{cyan}{#2}}
\newcommand{\vlcomment}[1]{\textcolor{Orange}{[vl: #1]}}

\newcommand{\dzadd}[1]{\textcolor{magenta}{#1}}
\newcommand{\dzremove}[1]{\textcolor{Gray}{\sout{#1}}}
\newcommand{\dzreplace}[2]{\textcolor{Gray}{\sout{#1}}\textcolor{magenta}{#2}}
\newcommand{\dzcomment}[1]{\textcolor{Orange}{[dz: #1]}}

\newcommand{\akadd}[1]{\textcolor{violet}{#1}}
\newcommand{\akremove}[1]{\textcolor{Gray}{\sout{#1}}}
\newcommand{\akreplace}[2]{\textcolor{Gray}{\sout{#1}}\textcolor{violet}{#2}}
\newcommand{\akcomment}[1]{\textcolor{Orange}{[ak: #1]}}

% To show changes in the document, comment out the \renewcommand definitions below
%\renewcommand{\vladd}[1]{#1}
%\renewcommand{\vlremove}[1]{}
%\renewcommand{\vlreplace}[2]{#2}
%\renewcommand{\vlcomment}[1]{}

%\renewcommand{\shadd}[1]{#1}
%\renewcommand{\shremove}[1]{}
%\renewcommand{\shreplace}[2]{#2}
%\renewcommand{\shcomment}[1]{}

%\renewcommand{\aladd}[1]{#1}
%\renewcommand{\alremove}[1]{}
%\renewcommand{\alreplace}[2]{#2}
%\renewcommand{\alcomment}[1]{}

%% ---------------------------------------
%% change this to hide/show the grumblers
%% and ``for our eyes only'' sections
%% ---------------------------------------
\newboolean{publicversion}
\setboolean{publicversion}{true}
%\setboolean{publicversion}{false}

\ifthenelse{\boolean{publicversion}}{
  \newcommand{\grumbler}[2]{}
  \newcommand{\new}[1]{#1}
}{
  \newcommand{\grumbler}[2]{\textcolor{blue}{\bf #1: #2}}
  \newcommand{\new}[1]{\textcolor{blue}{#1}}
}

% Add your own...

% Definitions for colors
\definecolor{heraldBlue}{rgb}{0.0,0.0,0.8}
\definecolor{heraldRed}{rgb}{0.8,0.0,0.0}
\definecolor{heraldGray}{rgb}{0.2,0.2,0.5}
\definecolor{heraldGreen}{rgb}{0.0,0.4,0.0}
\definecolor{heraldPink}{rgb}{0.8,0.1,0.6}

% Commands for common abbreviations
\newcommand*{\eg}{e.g.\@\xspace}
\newcommand*{\ie}{i.e.\@\xspace}

\makeatletter
\newcommand*{\etc}{%
    \@ifnextchar{.}%
        {etc}%
        {etc.\@\xspace}%
}
\makeatother

% Commands for author comments
\newcommand{\simon}[1]{\grumbler{SP}{#1}}
\newcommand{\hf}[1]{\grumbler{HF}{#1}}
\newcommand{\tsh}[1]{\grumbler{TH}{#1}}
\newcommand{\amanda}[1]{\grumbler{AR}{#1}}
\newcommand{\tim}[1]{\grumbler{TS}{#1}}
\newcommand{\mattan}[1]{\grumbler{ME}{#1}}
\newcommand{\mattanadd}[1]{\textcolor{purple}{#1}\xspace}
\newcommand{\mattancut}[2][]{#1}
\newcommand{\kayvan}[1]{\grumbler{KM}{#1}}

\thispagestyle{empty}
\pagestyle{empty}

\section{Introduction}\label{sec:intro}

Big Data applications increasingly look to tiered main memory to
increase memory capacity beyond machine-local DRAM with low
overhead~\cite{tmts}. High-capacity disaggregated memory such as via
Compute eXpress Link (CXL)~\cite{cxl} and non-volatile memory
(NVM)~\cite{optane} allow for systems with direct application access
to terabytes of main memory, without the overhead of swap-based (i.e.,
non-main) slow memory tiers. Still, both technologies provide lower performance, in terms of access latency and bandwidth,
compared to direct-attached DRAM. Hence, applications must tier data
among low-capacity fast memory and high-capacity slow memory to
achieve desired performance and cost.

% Applications want high and predictable performance to meet their 
% needs. Datacenter providers want to provide high and predictable 
% performance while minimizing their total costs, so they often 
% co-locate different applications on the same machine to increase 
% utilization of existing resources. Co-locating applications must be
% done carefully to avoid interference among applications, violating 
% quality-of-service promises.

To improve system utilization, application colocation on tiered memory
systems is desirable. Big Data applications have extremely large
memory footprints and migrating their state among servers is very
resource intensive. Hence, tiered memory systems for Big Data
applications seek to maximize colocation, but must also deliver
quality of service (QoS) among colocated processes and do so in
scenarios with dynamically changing workloads and memory
requirements. To minimize overhead that would otherwise negate any
benefit of increased utilization, tiering mechanisms must also be
lightweight. This is in opposition to existing approaches, which are
conservative in their allocation of slow memory~\cite{tmts}, have high
slow tier access overheads~\cite{tmo,software-far-mem}, or are
static~\cite{pond,hemem,optane}. % We explore Big Data application
% performance and server utilization implications of colocation without
% dynamic QoS (\S\ref{sec:tieredmem}) and show that tail latency and
% throughput can be affected by up to 35\% and 12\% versus isolation
% \simon{Up to date?}, respectively.

% In a multi-tenant environment with tiered memory, applications contend
% over the limited fast memory capacity. %  and slow memory bandwidth
% \amanda{Should we mention slow memory bandwidth? We state later on
%   that CXL bandwidth is config-dependent but can have similar
%   bandwidth to NVM read bandwidth, so it is still lower than local
%   DRAM and likely to be contended. We currently do not do any
%   partitioning of bandwidth, so I'm not sure this fits in right now}.

%A key challenge to the deployment of tiered memory in production
%datacenters and cloud environments is delivering quality of service (QoS). In a
%multi-tenant environment with tiered memory, applications may contend
%over limited fast memory capacity. To raise server utilization, the
%colocation of tasks is often inevitable. When colocated,
%latency-critical tasks are often impacted by low-priority tasks in
%unpredictable ways.

% QoS-aware tiered memory systems should provide balanced performance,
% while handling dynamically changing application demands. However,

We present \sys, a tiered main memory management system that improves
colocation performance for Big Data applications running on servers
with large slow memory tiers. \sys uses an application-agnostic and
lightweight memory occupancy control mechanism to provide QoS among
many Big Data applications sharing tiered main memory
simultaneously. \sys is designed to be easily modifiable and
extensible, without complex kernel code development. % As
% we will see, existing tiered memory systems do not adequately support
% this scenario~\ref{sec:tieredmem}.
Specifically, \sys uses these techniques:

\begin{compactitem}[\labelitemi]
\item \textbf{QoS-aware tiered memory management policy
    (\S\ref{sec:policy}).} We design a QoS-aware memory management
  policy based on \emph{fast-memory miss ratios} (FMMRs). FMMRs are a
  direct way to measure and specify tiered memory QoS for memory
  latency sensitive applications. % We describe how
  % task priorities can map to miss ratios if desired.

\item \textbf{Lightweight memory access sampling and binning
    (\S\ref{sec:mechanisms}).} Memory access frequency has a large
  impact on FMMRs. Efficient QoS support for multiple applications
  requires \sys to determine how frequently each part of each
  application's working set is accessed and how much of each working
  set to migrate to fast memory. \sys quickly and scalably identifies
  a \emph{memory heat gradient} for each tiered memory process by
  binning pages into heat groups. Based on QoS requirements and heat
  groups, \sys migrates the hottest and coldest pages among fast and
  slow memory to rapidly approach per-process target fast-memory miss
  ratios.

\item \textbf{Flexible user-space tiered memory management
    (\S\ref{sec:components}).} Inspired by user-space tiered memory
  management services, such as HeMem~\cite{hemem} and
  TMTS~\cite{tmts}, we design \sys to operate completely in
  user-space. This allows simple development and modification of
  policy and mechanisms without error-prone kernel
  development. To do so efficiently, we split \sys into a library and
  \emph{central manager} components (\S\ref{sec:components}), both
  running at user-level.

% \item \textbf{Avoiding fast memory bloating (\S\ref{sec:bloating}).}
%   Sparse hot sets increase the likelihood that hot data is co-located
%   with cold data, due to tiered memory management at page
%   granularity. This leads to a \emph{fast memory bloating problem},
%   where fast memory contains mostly cold data. As high-memory-capacity
%   applications often use huge pages, the problem is becoming more
%   severe. Fast memory bloating can cause performance interference
%   among applications, where a bloating application dominates fast
%   memory consumption.  \sys tracks when huge pages contain mixed data
%   and breaks them into constituent base pages when necessary. \sys
%   does this taking process priority into account.
\end{compactitem}

\noindent
We make the following contributions:

\begin{compactitem}[\labelitemi]
\item We study existing tiered main memory systems, detailing why they
  limit colocation of multiple applications
  (\S\ref{sec:background}).

\item We design (\S\ref{sec:design}) and implement
  (\S\ref{sec:impl}) \sys to improve colocation performance with
  dynamic Big Data application workloads on tiered main memory.

\item We evaluate and compare \sys to state-of-the-art hardware and
  software tiered main memory systems (\S\ref{sec:eval}). In
  particular, we compare \sys to static memory partitioning using
  HeMem~\cite{hemem}, as well as to automatic non-QoS management with
  Intel's hardware two-level memory caching (2LM)~\cite{optane} and
  Linux's AutoNUMA~\cite{autonuma}, on a system with tiered main
  memory comprising DRAM and Intel Optane persistent memory modules.
\end{compactitem}

\noindent
Our evaluation of the latency-sensitive FlexKVS~\cite{FlexNIC}
key-value store in several static and dynamic consolidation scenarios
with background workloads, such as GapBS~\cite{gapbs} and GUPS, shows
that \sys can uphold QoS in these scenarios. \sys always achieves the
configured target FMMR, providing up to 76\% and 80\% lower $90^{th}$
and $99^{th}$ percentile tail latency, while providing 11\% better
throughput for FlexKVS in consolidated workload mixes compared to
HeMem. Compared to AutoNUMA, \sys provides up to an order of magnitude
lower $90^{th}$ and $99^{th}$ percentile tail latency, while providing
38\% better throughput. \sys provides QoS with dynamically changing
workloads, while HeMem cannot provide dynamic QoS and AutoNUMA cannot
provide QoS at all.

%%% Local Variables:
%%% mode: latex
%%% TeX-master: "paper"

% !TeX root = paper.tex

\section{Background}\label{sec:background}

Tiered main memory enables applications access to memory capacities
beyond the reach of direct-attached DRAM. To do so, commercially
available tiered memory systems are configured using two memory tiers:
\emph{fast memory} and \emph{slow memory}. In tiered memory systems,
the fast memory tier offers lower latency and (typically) higher
bandwidth than the higher-capacity slow memory tier. Fast memory is
currently provided by direct-attached DRAM, while slow memory is
currently provided by high-capacity NVM (e.g., via Intel's
Optane~\cite{optane} persistent memory modules). CXL-based
disaggregated memory is expected to be offered as another option for
the slow memory tier (e.g., via Samsung's CXL memory
expanders~\cite{cxl_expander}).
% \paragraph{NVM versus CXL as slow memory tier}
We use Optane for the slow memory tier in this paper, as it is the
only retail option. However, we intentionally ignore NVM-specific
performance effects (e.g., read/write asymmetry and access
granularity), as we target tiered memory generally and we expect our
results to generalize to CXL-based slow memory\footnote{CXL
  performance in terms of unloaded latency is expected to be similar
  to that of Optane~\cite{cxl,pond}. Regarding bandwidth, Microsoft
  reports a CXL-based slow memory configuration~\cite{pond} that is
  comparable to the read bandwidth available in our evaluation
  platform (\S\ref{sec:eval})~\cite{hemem}.}.
% Optane's read bandwidth in a fully populated system~\cite{hemem}. Our
% evaluation platform (\S\ref{sec:eval}) is also fully populated with
% NVM.
% We expect tiered
% memory with Optane to have good predictive power to tiered memory
% performance with CXL.

% Tiered memory systems for Big Data applications need to maximize
% utilization on consolidated servers, while handling dynamically
% changing application demands. % Today's tiered memory systems' QoS
% support is insufficient. 
We characterize the performance properties of Big Data workloads and
motivate colocation on tiered memory servers (\S\ref{sec:workloads}). % We then describe how workloads are typically
% colocated in datacenter environments (\S\ref{sec:colocation}). 
We then describe why current tiered memory systems limit Big Data
application performance and colocation
(\S\ref{sec:tieredmem}).%  Finally, we analyze the application-level
% performance impact of tiered memory interference due to colocation, in
% terms of latency and throughput, quantifying why existing approaches
% fall short (\S\ref{sec:interference}).

% x Background on datacenter workloads and colocation?
% x "Latency-critical" workloads vs. "best effort" workloads
% x QoS and QoS metrics

\subsection{Big Data Workload Colocation}\label{sec:workloads}

Modern applications are increasingly memory-intensive. Web
applications access large in-memory key-value stores to build dynamic
web pages \cite{facebook_photo}, graph processing systems scour large
in-memory datasets to quickly answer analytical questions
\cite{graph500}, and machine learning systems train on huge in-memory
datasets to maximize a reward function \cite{parameterserver}. These
applications have large memory capacity requirements. Further,
\emph{latency-sensitive} (LS) applications have response latency
requirements, while \emph{best-effort} (BE) ones do not.

Dedicating servers to individual Big Data applications risks
underutilizing expensive memory and compute capacity and is
wasteful. Hence, and due to their varying requirements and behaviors,
LS and BE tasks are often colocated on servers~\cite{borg}. This
allows operators to increase server utilization by running BE tasks
during periods of low LS task utilization. To continue to provide QoS
to LS tasks, the operating system is tasked to promptly reallocate
potentially oversubscribed resources when an LS burst arrives. For
CPU, network, and storage, this involves context switching and deficit
round robin scheduling of requests to these resources, given task
priorities. For memory, existing approaches may kill lower priority
tasks to free memory when high priority tasks require
it~\cite{autopilot} or migrate tasks to machines with appropriate
resources~\cite{tmts}.

% \simon{Describe memory bandwidth as a QoS issue?}

Killing or migrating Big Data applications is undesirable. Big Data
applications have extremely large memory footprints and recreating or
migrating their state is very resource intensive. Instead, we wish to
improve server utilization by dynamically managing available tiered
main memory, aggressively allocating from both memory tiers, while
providing lightweight memory access. This is in contrast to existing
approaches, which are conservative in their allocation of slow
memory~\cite{tmts}, have high slow tier access
overheads~\cite{tmo,software-far-mem}, or are
static~\cite{pond,hemem,optane}. Given the burstiness of LS tasks,
these approaches lead to the underutilization of tiered memory and a
hard limit on the number of applications that may be simultaneously
scheduled for execution on a
machine. % As a result, memory performance will be worse than

\subsection{Existing Tiered Memory Systems}\label{sec:tieredmem}

In this paper, we explore the design of tiered main memory management
for Big Data applications that do not fit into the fast tier of a
single server. Big Data applications require us to focus on
lightweight mechanisms and policies that improve application
colocation and performance with a large slow tier, rather than
focusing on slowdown versus the fast tier. Big Data applications also
cannot be easily migrated to other servers due to their large state,
limiting the opportunity of distributed mechanisms. In this section,
we study existing tiered memory systems and whether they achieve these
goals.

TMTS~\cite{tmts} makes the case for tiered main memory system
deployment at scale. TMTS showcases significant cost saving
opportunities even when offloading only a fraction of cold data to a
slow memory tier. It also demonstrates that lightweight tiered memory
management mechanisms are necessary to achieve these benefits for
tiered main memory. Focusing on applications with memory footprints
that fit in the fast memory tier, TMTS describes the design of a
distributed tiered memory management system that achieves these
benefits by avoiding SLO impact for these applications versus baseline
execution in the fast memory tier. However, to stay within strict
performance bounds versus the fast memory tier, TMTS must be
conservative, limiting slow tier capacity to 25\% of the fast tier, as
well as keeping any accessed data in the fast tier. This requires TMTS
to use conservative page scanning techniques in addition to memory
sampling, slowing memory management timescales to $O(minutes)$. TMTS'
focus on avoiding SLO impact versus the fast memory tier limits
colocation and slow tier capacity scalability that are necessary to
support Big Data applications.

Other existing systems, such as TPP~\cite{tpp}, Optane two-level
memory (2LM)\footnote{2LM is identical to ``Memory Mode'' (e.g.,
  described in Section 2.4 in \cite{hemem}) for Intel's Optane
  persistent memory, but it also supports CXL.}~\cite{optane},
Pond~\cite{pond}, AutoNUMA~\cite{autonuma} with CPU-less NUMA
nodes~\cite{cpuless_numa}, and HeMem~\cite{hemem} are also designed
for tiered main memory. However, they do not provide dynamic QoS. 2LM
and AutoNUMA cannot differentiate between applications accessing
tiered memory, leading to performance interference. HeMem is a
per-process tiered memory library. To support multiple processes, fast
and slow memory have to be statically partitioned and individually
managed by separate HeMem instances. Pond~\cite{pond} relies on
offline working set estimation. Static partitioning can provide QoS
when partitions can be sized according to application requirements and
workloads do not change. In practice, workloads often vary dynamically
and unpredictably. Dynamic QoS is hence necessary to enable dynamic
Big Data application colocation. Without it, colocation opportunities
are reduced to preallocated memory areas or time slots.

Some systems, like TMO~\cite{tmo} and Software-Defined Far Memory
(SFM)~\cite{software-far-mem} provide dynamic QoS, but are designed
for swap-based slow memory (SSDs and compressed DRAM). Their policies
and mechanisms rely on heavy-weight page faults and IO to measure and
manage application tiered memory access. These heavyweight mechanisms
drastically reduce access performance to the slow memory tier,
prohibiting its frequent use by Big Data applications.

\section{\sys Design}\label{sec:design}

\sys is a user-space tiered memory management system that can
dynamically manage multiple processes with different QoS
requirements. \sys has a number of design goals:

\begin{compactitem}[\labelitemi]
\item \textbf{Simplicity.} Setting of tiered memory QoS goals should
  be simple and intuitive. In \sys, users/controllers configure a
  per-process target FMMR $t_{miss} \in [0, 1]$, representing how
  often that process accesses data from the slow memory tier as a
  fraction of all memory accesses. The FMMR is a direct way to specify
  per-process memory system QoS. It is easy to assess to what extent
  the goal is met by the system. It is therefore also a good target
  for higher-level resource-management systems because it directly and
  monotonically correlates with performance (e.g.,~\cite{heracles}).

\item \textbf{Scalability.} Tiered memory is expected to scale to
  terabytes and modern cloud servers are expected to multiplex tens of
  applications. \sys manages these memory sizes across processes by
  sampling memory accesses, binning pages into heat classes, and
  determining how to re-allocate memory across processes efficiently.
  Sampling overhead grows with memory bandwidth, scaling more slowly
  than memory capacity and core count.

\item \textbf{Flexibility.} Tiered memory management is an evolving
  domain and it should be simple to extend and modify the management
  system. \sys is flexible by managing tiered memory at user-level. To
  do so, \sys consists of two main components: a \emph{central
    manager} process and a library component (\libsys) that is
  dynamically linked into unmodified processes wishing to use tiered
  memory.
\end{compactitem}

In this section, we describe the simple \sys policy
(\S\ref{sec:policy}), lightweight mechanisms that scalably inform the
policy (\S\ref{sec:mechanisms}), as well as \sys's flexible user-space
design (\S\ref{sec:components}). Finally, we discuss how several
tiered memory management issues may be handled by \sys
(\S\ref{sec:discussion}).

Figure~\ref{fig:qos_design} shows an overview of \sys's design as it
relates to providing QoS for tiered memory. The figure illustrates in
a concrete example, where two processes with target FMMR 0.9 and 0.1
are managed by \sys, the execution at the end of a \sys policy epoch
(\S\ref{sec:policy}). The following sections refer to the figure,
where appropriate.

\subsection{Tiered Memory QoS Policy}\label{sec:policy}

\begin{figure}
  \centering
  \includegraphics[width=\columnwidth]{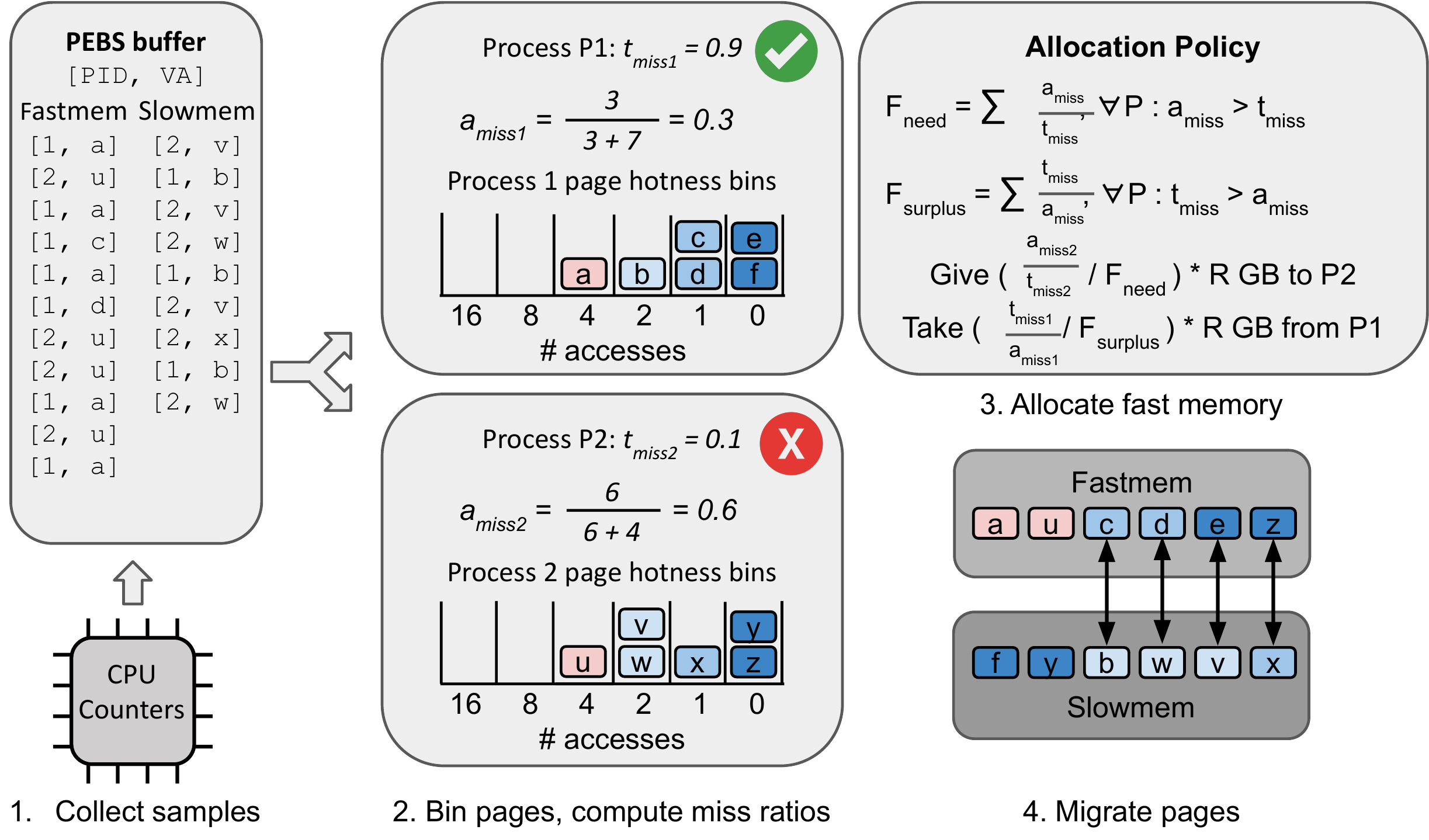}
  \caption{\sys QoS policy example.}
  \label{fig:qos_design}
\end{figure}

How often a memory-intensive application accesses the slow tier has a
direct effect on overall application performance~\cite{tmts}. The goal
of \sys's tiered memory QoS policy is thus to allocate the limited
fast memory tier such that each process is able to meet its configured
$t_{miss}$. %Processes with lower $t_{miss}$ are given higher priority
%than processes with higher $t_{miss}$.
%
Users configure a per-process target FMMR $t_{miss} \in ]0, 1]$,
representing how often that process accesses data from the slow memory
tier as a fraction of all memory accesses. The FMMR is a direct way of
specifying a per-process memory system QoS target. An FMMR of 1
implies no QoS and \sys may place the entire process working set into
slow memory if necessary to provide QoS to other processes. To obtain
an FMMR of 0, tiered memory should simply be disabled for that
process.

To provide QoS, \sys samples per-process memory accesses to fast
($a_{fast}$) and slow memory ($a_{slow}$), as described in
\S\ref{sec:mechanisms} and shown in step 1 of
Figure~\ref{fig:qos_design}. \sys then computes each process's current
FMMR $a_{miss} = \frac{a_{slow}}{a_{slow} + a_{fast}} \in [0, 1]$, as
shown in step 2 of Figure~\ref{fig:qos_design}. We assess $a_{miss}$ 
as an exponentially moving average (EWMA) with $\lambda = 0.5$ every
second---the \sys policy epoch duration
(cf.~\S\ref{sec:sensitivity}). If $a_{slow} = a_{fast} = 0$, we set
$a_{miss} := 0$ for that epoch. Hence, memory-inactive processes will
eventually give up their fast memory allocation.

To meet FMMR targets, \sys needs to accomplish two goals. First, fast
memory must be partitioned among processes such that each process has
the fast memory it needs to meet its FMMR target. Second, since fast
memory is a constrained resource in our system, where slow memory is
many times larger than the fast memory, \sys must ensure that fast
memory is optimally utilized by keeping each application's hottest
pages in fast memory. Both of these goals require \sys to migrate
pages. We cap \sys's migration rate to 4 GB per epoch. %, identical to
%HeMem's migration rate~\cite{hemem}. 
\sys allocates half of the
migration rate to each goal.

\paragraph{Fast memory reallocation.} It is impossible to determine
the ideal fast memory partition a priori, as it depends on future
access patterns. Instead, \sys reallocates fast memory proportionally
among processes, based on distance from $t_{miss}$. To do so, \sys
first calculates the total scale of needed and surplus fast memory,
$F_{need}$ and $F_{surplus}$ w.r.t.\ $t_{miss}$ (as defined in
Figure~\ref{fig:qos_design}). These are dimensionless quantities, used
to calculate a fractional migration bandwidth $M_p$ for each process $p$
in the epoch. 
Fast memory
$M_p = \frac{a_{miss}}{t_{miss}} \times \frac{R}{F_{need}}$ is given
to processes $p$ with $a_{miss} > t_{miss}$.
%
% $$F_{give} = \sum{\frac{t_{miss}}{a_{miss}}}, \forall P : t_{miss} > a_{miss}$$
% $$F_{need} = \sum{\frac{a_{miss}}{t_{miss}}}, \forall P : a_{miss} > t_{miss}$$
%
Fast memory
$M_p = \frac{t_{miss}}{a_{miss}} \times \frac{R}{F_{surplus}}$ is
taken from processes $p$ with $a_{miss} < t_{miss}$ if they have fast memory. 
R is the total migration bandwidth \sys uses for fast memory
reallocations each epoch. If $M_p$ is larger than the amount of fast
memory available to $p$, then \sys simply takes all of its remaining
fast memory. In this case, \sys may underutilize its migration rate
within an epoch. When $t_{miss} = a_{miss}$, the process has the
appropriate amount of fast memory and we maintain its
allocation. In calculations, where $a_{miss}$ is a 0 denominator, we
substitute $\infty$ for the result and we define
$\frac{\infty}{\infty} = 1$. If there are multiple processes with
$a_{miss} = 0$, \sys takes fast memory from only one per epoch.

It is possible for \sys to encounter an application mix where it is 
impossible to meet each application's target FMMR due to not having enough
fast memory available. In this case, \sys attempts to meet the target FMMR
for as many applications as it can, on a first-come-first-served basis. \sys
can flag the applications that cannot meet their target FMMRs so that a system
administrator can decide whether to wait until more fast memory becomes
available (due to, e.g., process exit) or to kill that application and move it
to a machine with adequate fast memory. An alternative policy might attempt to 
provide fairness in these situations by allocating fast memory such that each
application is equally close to their target. Note that fast memory
reallocations entail data movement overheads and can affect application
performance, particularly with bandwidth-constrained slow tiers. Thus, \sys
minimizes fast memory reallocations in situations where it cannot meet all
target FMMRs by stopping once it has met all the target FMMRs it can

\paragraph{Page migration.} Once fast memory allocations are decided
for each process, \sys fills the allocations by migrating hot pages to
them, while colder pages previously residing in fast memory are
migrated to slow memory. Note that this step occurs for each process
regardless of whether its fast memory allocation changed. This
ensures, even in the absence of fast memory reallocations, that each
process's hot data resides in fast memory, while cold data resides in
slow memory.

To maximize the impact that migrations have on fast memory miss
ratios, \sys migrates each process's hottest pages to fast memory and
coldest pages to slow memory. To do so, \sys uses a memory heat
gradient computed via per-process hotness bins. We describe this
process in \S\ref{sec:mechanisms}. Step 4 of
Figure~\ref{fig:qos_design} shows the hottest pages x, v, and w of
process 2, as well as page b of process 1 are migrated to fast memory,
while the coldest pages c, d, and e of process 1, as well as page z of
process 2 are migrated to slow memory. The figure only shows this
subset of page migrations, while more pages would be migrated to
fulfill the fast memory reallocation. As migration is carried out
for all processes, process 1 migrates page b to fast memory, even
though its fast memory allocation shrank.

\paragraph{Memory allocation.} When physical memory is allocated to a
process, either explicitly via \verb+mmap+ with the
\verb+MAP_POPULATE+ flag, or implicitly via a page fault, \sys first
attempts to allocate fast memory. If no fast memory is available to
the process, then slow memory is allocated. If slow memory is also
unavailable, then an error is returned (for \verb+mmap+) or the
process is killed (for page faults).

% \simon{How are page faults for fresh memory handled?}
% \amanda{Currently in a similar manner to HeMem. If there are free fast
%   memory pages, then the page fault can allocate pages from fast
%   memory. Otherwise it handles it from slow memory. Could maybe do
%   something with priorities here too? But then we might have
%   migrations on the fault path, but since faults wouldn't happen in
%   steady-state anyway maybe it is fine?}\simon{Not sure your answer is
%   correct, since you say above that each process is given a starting
%   allocation of fast memory. In that case, it doesn't work like
%   HeMem. Instead, each process only gets a limited number of fast
%   memory pages, rather than checking all free fast memory pages. Can
%   you comment?} \amanda{Right, sorry. The fault can take a free DRAM
%   page if the application's allocation of DRAM allows it to, otherwise
%   it takes a free NVM page.}

\paragraph{Process exit.}
% Each process is given a starting allocation of fast memory on
% startup. An initial allocation of fast memory ensures that each
% process a high priority process is able to use some fast memory right
% away, even if hotness and fast memory miss rates have not yet been
% determined. \simon{How large is this starting allocation?}
%
% If only one process has joined, then its initial allocation of fast
% memory is equal to the size of fast memory. If one or more processes
% is already running when a new process joins, fast memory may need to
% be reallocated from the currently running processes. This is done
% according to the QoS policy, described above.
When a process exits, \sys reclaims its memory and returns it to the
free memory pool. If, at this point, processes exist with
$a_{miss} > t_{miss}$, then memory is immediately allocated from the
free memory pool.

% \sys simply divides up the newly freed fast memory evenly among the
% remaining processes. \amanda{probably can come up with someting a
% bit smarter here}\simon{Shouldn't it be simply returned to the free
% pool?}

\subsection{Lightweight QoS Mechanisms}\label{sec:mechanisms}

\sys relies on a number of mechanisms to inform its QoS policy
(\S\ref{sec:policy}). The ability of any process to achieve a certain
FMMR is determined primarily by how frequently memory is
accessed. Many applications have non-uniform memory access
patterns. Hence, \sys tracks per-page memory access frequencies of
each process. \sys uses memory access sampling to track page access
frequencies and then categorizes pages into different hotness bins,
based on how frequently each page is being accessed.

\paragraph{FMMR sampling.}
\sys monitors per-process FMMR via light-weight sampling of memory
access patterns (cf. HeMem~\cite{hemem,tmts}). Two performance
counters are configured to separately sample all cache misses served from
fast and slow memory and record the virtual address target of each
such instruction as well as the PID of the process the instruction is
from (\S\ref{sec:impl}). This is shown in step 1 of
Figure~\ref{fig:qos_design}. \sys uses a sampling period of 100 load
events (or 1\% of loads), which provides adequate sampling
fidelity~\cite{tmts}. Each epoch, \sys computes a current FMMR for
each process it is managing, based on the samples from fast and slow
memory during that epoch. Sampling is a scalable mechanism as it does
not depend on memory capacity and its overhead scales with memory
bandwidth, which is outpaced by overall performance and core-count
scaling.

\paragraph{Hotness bins.}
\sys also uses the samples to determine access frequencies for each
page of memory in each managed process. Each epoch, after quantizing
samples to huge page granularity, \sys accumulates accesses per
observed page. \sys then categorizes each process's pages into a
configurable number of per-process \emph{hotness bins}, based on
accumulated accesses. Bins represent exponential access frequencies,
relative to each other---a page in one bin has been accessed roughly
twice as much as a page in its neighboring colder bin; the coldest bin
represents pages that have not been recently accessed. Example hotness
bins of processes 1 and 2 are shown in step 2 of
Figure~\ref{fig:qos_design}.

Once enough samples accumulate to a page, the page is promoted to its
hotter neighbor bin. We configure 6 per-process hotness bins, which we
empirically determined offer good fidelity for QoS policy
decisions. Once a page reaches the threshold for promotion beyond the
maximum of the configured bin (cf.\ $2^5$ in our 6-bin configuration),
\sys ``cools'' all pages by halving all sample counts, rounded down,
redistributing each page to a one-cooler bin and leaving the hottest
page (momentarily) alone in the hottest bin. Cooling happens at most
once per epoch.
%as well as a \emph{hotness threshold} of 32 page
%access samples. Once any page accumulates 32 samples, \sys  A hotness threshold of 32 with 6
%bins forces the coldest bin to hold all pages without any accumulated 
%samples.

Our implementation of the hotness bins is simple and cheap. Each bin
requires a single counter for the number of pages in that bin and two
linked-list head pointers for associating pages in that bin from fast
and slow memory. Victims for promotion to fast memory and demotion to
slow memory can thus be quickly identified with low overhead. A
counter is also stored per page for its accumulated accesses. This
counter and the bins themselves are lazily updated when a page
receives a sample or when it is considered for migration; this
includes applying the accumulated cooling events since its previous
counter update. While this allows essentially free scaling 
of the number of hotness bins, a larger number of bins increases the
cooling interval and slows down the response to hot-set changes.
% \mattan{This feels like it belongs here, but can also be moved to the
%   implementation section.}
%\kayvan{Is this okay? Hopefully this captures the two obvious difficulties.}
%The implementation of the hotness bins can be divided into 2 parts:
%updating the bins, and victim selection. The hotness bins are updated lazily
%meaning that pages only move between bins when they recieve a sample
%or when they are considered for migration. This keeps the cost of
%maintaining the bins low. Selecting a victim from the hotness bins
%requires looping over the bins to determine if there are any pages
%in that bin. This requires only loop one iteration per bin, with
%each iteration checking the bin's current length.
%Hotness bins have negligible computational overhead so a higher number
%is desirable for higher fidelity. We found that 6 bins are enough for
%our experiments, which involved up to 6 processes running concurrently.
%% referencing number of bins
%%\amanda{Kayvan: why 6?}  \simon{Sensitivity analysis in eval?}

% referencing A hotness threshold of 32 with 6
% bins forces the coldest bin to hold all pages without any accumulated 
% samples 
%\simon{Significant?} \amanda{Not sure what you mean here.}
%\simon{Does it make a difference? Could the coldest bin also hold
%  pages with some accesses and \sys performance would be the same?}.

\sys uses the hotness bins to determine a \emph{memory heat gradient}
per process. The memory heat gradient allows \sys to quickly determine
which pages have the greatest impact on a process's FMMR. For each
process, after determining its fast memory allocation for the epoch,
\sys migrates pages to fast memory starting with the hottest bin. To
make room, \sys evicts pages to slow memory starting with the coldest
bin.

\subsection{Userspace Tiered Memory Management}\label{sec:components}

\begin{figure}[tb]
  \centering
  \includegraphics[width=\columnwidth]{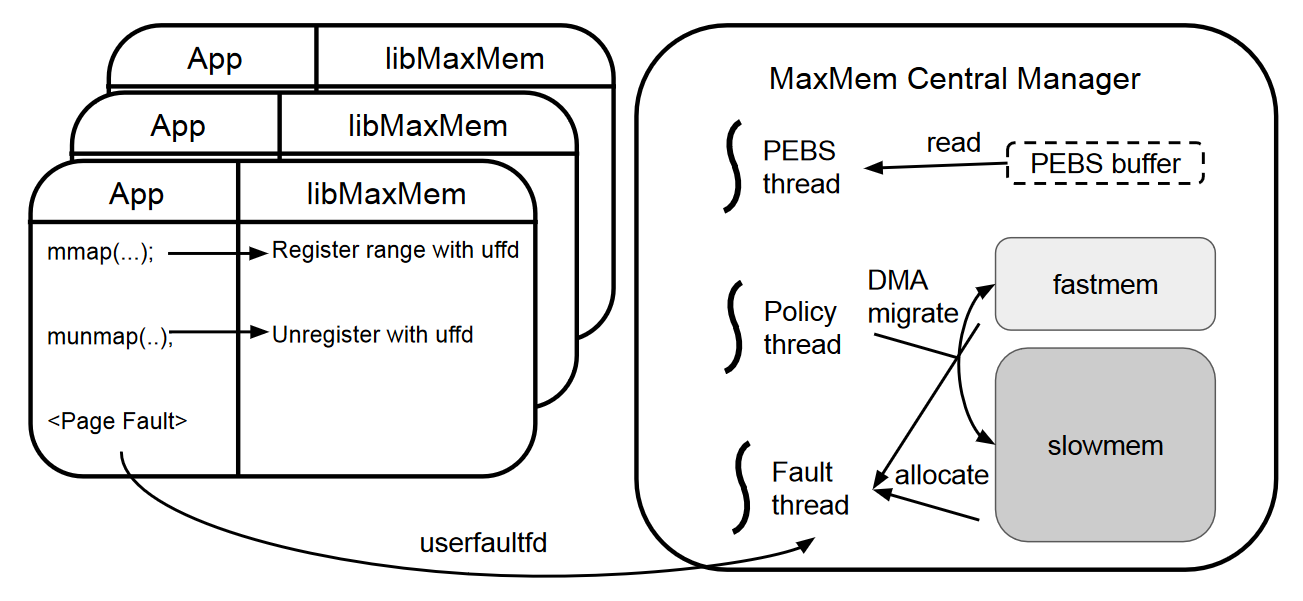}
  \caption{\sys user-space design.}
  \label{fig:central-manager-design}
  % \end{wrapfigure}
\end{figure}

To manage tiered memory efficiently from user space, \sys is split
into two components that each run at user-level: a library component
(\libsys) that is linked into processes and a central manager that
runs as a separate process. An overview of this design is given in
Figure~\ref{fig:central-manager-design}.

The central manager is in charge of allocating memory to processes
from fast and slow tiers, as well as migrating process memory between
tiers based on patterns gathered from memory access samples. \libsys
is responsible for intercepting application virtual memory allocation
calls and registering the corresponding memory regions with the
central manager for tiered memory management. To do so, \libsys
registers regions with a user-level page fault handler (via
userfaultfd~\cite{userfaultfd}) executing in the central
manager. Userfaultfd has negligible overhead for Big Data
applications~\cite{hemem}. Processes do not control their tiered
memory allocations.

\setlength{\columnsep}{5pt}

\paragraph{Central manager.} The central manager runs as a separate
daemon process and listens on a UNIX domain socket for new \libsys
connections. Tiered memory processes connect to the socket on startup
via \libsys and send a userfaultfd file descriptor. A separate fault
handling thread monitors these descriptors for page faults. Upon a
page missing fault, the central manager allocates a page to the
faulting process based on its policy (\S\ref{sec:policy}). Like
HeMem~\cite{hemem}, \sys write-protects pages as it migrates them.
When a protection fault occurs on a page under migration, the central
manager waits until the migration completes before un-protecting the
page.

\paragraph{PEBS thread.}
The central manager uses Intel processor event based sampling
(PEBS)~\cite{pebs} for light-weight sampling of per-process memory
access patterns (cf.~HeMem~\cite{hemem}). A separate PEBS thread
monitors the PEBS buffer and updates page access information.

\paragraph{Policy thread.}
A policy thread implements the QoS policy and runs once every
epoch. The policy thread uses $t_{miss}$ and page access statistics
from the PEBS thread to make decisions about which pages should be
placed in which memory tier. % The policy thread enqueues a list of
% pages for migration to the migration thread.
%
% \paragraph{Migration thread}
When migrating pages, the policy thread first write protects each page
to ensure correctness before copying the page's data to its new
location. The process may read from the page while it is migrating but
any write results in a write-protection fault that is caught by the
fault thread. When migration is finished, the policy thread resets
page protection bits.

\paragraph{\libsys.} \libsys is simple. Upon startup, \libsys opens a
UNIX domain socket to the central manager. It then opens a userfaultfd
file descriptor and sends it to the central manager. \libsys then
intercepts the \verb+mmap+ memory allocation system call. Large
(greater than 1~GB) allocations are registered with userfaultfd,
allowing the central manager to manage the memory backing these
allocations. Any page fault from the process on these regions is
forwarded by userfaultfd to the central manager, which is responsible
for allocating memory to handle the fault. When a process frees
memory, \libsys removes the region from userfaultfd, which notifies
the central manager of the freed memory.

% \subsection{Avoiding Fast Memory Bloating}\label{sec:bloating}

% \amanda{Will we include densely vs. sparsely hot huge pages?}
% \amanda{Currently seems unlikely we will get to this before OSDI...}

\subsection{Discussion}\label{sec:discussion}

\paragraph{Application weights.} As $t_{miss}$ of an application is
decreased, its share of fast memory and its performance
increase. Different weights can thus be expressed as relative values
of FMMRs. Resource-provisioning and QoS runtimes
(e.g.,~\cite{heracles}) can hence directly use $t_{miss}$ because it
is monotonic with performance. We argue that using relative miss
ratios is preferable to directly provisioning fast memory because the
application miss curves are not known a-priori.
%\sys's target miss ratios can be
%mapped to relative application priorities for memory access
%performance. For example, if one application has double the priority
%of another, the first application can be given a target miss ratio
%that is half the target miss ratio of the second application. \sys
%then prioritizes the first application for fast memory when
%capacity is limited.
  % \mattan{I changed this text and left the original commented out in TeX. Simon's note below refers to the previous text, but I left it here to see if my suggested text addresses the concern.}\simon{Too simplistic. Better to provide a
  % formula or some other more direct means than this example to show
  % how to convert between the two. Note that priorities are not
  % weights. There is no real notion of ``double priority'' like there
  % would be with weights.}

% , making it more likely that this higher priority
% process has the memory resources available to enable it to meet its
% other SLOs. \simon{That's not the same as process priorities. Focusing it
% just on memory.}
%   \simon{Didn't Rajath point out a corner case in 590s
  % where a high priority app could still receive less fast memory than
  % a low priority one, hence inverting priorities?} \amanda{Does this
  % help clarify?}

\paragraph{Application performance metrics.}
\sys uses FMMR to gauge the effect of slow memory on application
performance. Other such metrics have been
proposed. SFM~\cite{software-far-mem} defines a \emph{promotion rate}
at which pages are swapped in from the slow memory tier and seeks to
keep the promotion rate for an application below a target 0.2\% of the
application's working set per minute. In contrast, \sys's FMRR can
vary for each application and is measured over seconds, allowing \sys
to provide different levels of QoS to applications and to react over
shorter timescales. TMO~\cite{tmo} uses \emph{pressure stall
  information (PSI)} to quantify lost work due to a shortage of
resources---a holistic view of application QoS. PSI could be
integrated with \sys to derive FMMRs if tiered memory is the main QoS
factor.

%\paragraph{Thrashing.} In a system with multiple processes competing
%for fast memory, thrashing could occur where fast memory pages are
%migrated back and forth between two processes. \sys's policy prevents
%this thrashing by reallocating fast memory from a process with a lower
%target miss ratio to a process with a higher target miss ratio if the
%first process is at or lower than its target miss ratio.  If both
%processes are not meeting their target miss ratio, then only the
%process with the lower target miss ratio would be allocated more fast
%memory at the expense of the process with the higher target miss
%ratio. \simon{This is oddly specific and doesn't answer the thrashing
%question. What if the processes constantly hit and then miss their
%ratio? Wouldn't that cause thrashing?}

\paragraph{Fair sharing.} \sys's fast memory allocation policy is
best-effort based on application target FMMRs. \sys attempts to
allocate enough fast memory to each application to allow it to meet
its target. If more fast memory is still available at this point, then
\sys allocates the remaining equally to all processes. Fast memory
reallocations are done proportionally to how close or far each
application is from its target. Applications substantially below their
target FMMRs give up fast memory faster than applications slightly
below their target FMMRs. Similarly, applications substantially above
their target FMMRs are allocated fast memory faster than applications
slightly above their target FMMRs.

\paragraph{Containers and virtual machines (VMs).} \sys's policies and
mechanisms naturally extend to containers and VMs. Fast memory miss
ratios and page access frequencies may be specified and measured on a
per-container and per-VM basis. Linux's performance monitoring
infrastructure, which \sys uses, supports monitoring containers based
on control groups (cgroups) via the \verb+PERF_FLAG_PID_CGROUP+
flag. VMs may be identified simply via PIDs and host execution may be
excluded from monitoring via the \verb+exclude_host+ flag.

\paragraph{Security.} We have designed \sys for a standard OS threat
model, where user-space processes can request service and OS
processes, including the \sys central manager, are trusted to provide
it globally. Hence, mechanisms and policies for memory allocation,
measurement, and migration are implemented in the central manager,
enabling realization of \sys's QoS management policy without
interference from user-space processes (with one
implementation-specific caveat described in \S\ref{sec:impl}). As is
typical of best-effort OS policies, it can be abused. For example, a
malicious process can continually touch any unused memory ranges to
cause \sys to migrate them to the fast memory tier and inflate the
process's level of service. To remedy abuse, the system administrator
can explicitly limit memory allocation.

\paragraph{Starvation.} Resource managers that can deprive tasks of a
particular resource may prevent the task from making progress, also
known as \emph{starvation}. In \sys, it is possible that tasks are
deprived of fast memory. In this case, the tasks are still able to
access slow memory and make progress, so starvation is not an issue in
\sys.

\section{Implementation}\label{sec:impl}

We implement \sys in around 4,500 lines of code for the central
manager process and around 750 lines for the \libsys component linked
in each application. The \sys central manager reuses mechanisms from
the HeMem~\cite{hemem} tiered memory management system. In particular,
PEBS sampling, DMA migration, and userfaultfd page fault handling.

Our prototype expects slow and fast memory to be exposed as DAX files
to user-space. For example, Intel Optane's App-Direct mode exposes a
DAX file for slow memory. We can expose direct-attached DRAM as a DAX
for fast memory at boot via the Linux kernel \verb+memmap+
command-line argument. Dynamically sharing DRAM with the existing
kernel memory manager is possible, for example via pinned anonymous
mappings, but requires additional userfaultfd support for these
mappings~\cite{hemem}.

\paragraph{Memory access sampling.} The central manager uses the
\verb+perf_event_open+ system call to set up performance monitoring
counters to monitor per-process memory access patterns with PEBS. The
PEBS counters are set up to sample from all loads that are served from
DRAM\footnote{\texttt{MEM\_LOAD\_L3\_MISS\_RETIRED.LOCAL\_DRAM}} and
NVM.\footnote{\texttt{MEM\_LOAD\_RETIRED.LOCAL\_PMM}} \sys configures
each core with its own set of PEBS counters configured to count any
events occurring on that core. In addition to the virtual memory
address target of the instruction, the PID of the process producing
the sample is included so that \sys can distinguish samples from the
different processes it is managing.

\paragraph{Memory migration.} The central manager migrates memory
among tiers with a DMA engine~\cite{ioat} if available. We implement a Linux kernel driver that exposes a DMA API to
user-space, which is used by the central manager. If a DMA engine is
not available, the copy is performed by 4 parallel copy threads, as
done in HeMem~\cite{hemem}.

\paragraph{Memory mapping.} The Linux kernel has only limited support
for modifying another process's virtual address space. While mapping
memory in response to page faults is possible via userfaultfd, there
is no support to remap another process's mapped memory. Since
remapping is necessary to support memory migration, our \sys prototype
realizes memory remapping from within the process. We do so by
remapping within \libsys, cooperatively, upon request from the central
manager. The central manager sends a message to a process when serving
a page fault or when migrating a page to inform the process where the
page of memory should be mapped and \libsys carries out the
mapping. While doing so is not safe for production deployment, it lets
us avoid a potentially cumbersome implementation within the Linux
kernel, without adverse effect on our performance results. userfaultfd
could be extended to support this functionality.

\section{Evaluation}\label{sec:eval}

We first analyze the performance of \sys's policies and mechanisms
with a number of GUPS microbenchmarks (\S\ref{sec:microbench}). We
then evaluate a number of big data applications in a variety of
different colocation configurations (\S\ref{sec:app_eval}). These big
data applications include a key-value store, a graph processing
system, and a parallel application
benchmark. % \amanda{and another app?
  % Something from NAS if it will work out of the box with
  % \sys?}\simon{Could try Silo with a workload that has
  % locality. TPC-H?}
Experiments are repeated five times and we report the
average. Timelines are from representative runs. There is negligible
variance across runs.

Our evaluation seeks to answer the following questions:

\begin{compactitem}[\labelitemi]
\item What are the overheads of the various policies and mechanisms of
  \sys? Can \sys offer competitive performance to other tiered memory
  management systems with only a single application running?
  (\S\ref{sec:overheads})

\item How well does \sys perform under a number of different and
  changing colocation scenarios? Can \sys accurately recognize when an
  LS application needs more DRAM to achieve its performance goals? Can
  \sys provide fair performance to applications in the same priority
  class? Can \sys adapt to dynamically changing access patterns and
  QoS requirements? (\S\ref{sec:performance})

\item How well does \sys meet the performance goals of an LS
  application in terms of tail latency and throughput?
  (\S\ref{sec:app_eval})

\item What is \sys's performance sensitivity to its various
  measurement and migration parameters? (\S\ref{sec:sensitivity})
\end{compactitem}

\paragraph{Evaluation platform.} We run our evaluation on a single
socket\footnote{NUMA effects of NVM are complex and have, in the
  context of storage, been analyzed before~\cite{odinfs,assise}. They
  are beyond the scope of this work.} of a dual-socket Intel Cascade
Lake-SP system running at 2.2GHz with 24 cores per
socket. Hyperthreads are disabled. Each socket has 192 GB of DDR4 DRAM
and 768 GB of Intel Optane DC NVM. There are 6 DIMMs of DRAM and NVM
per socket, fully leveraging all 6 memory channels. The machine runs
Debian 10.9 with Linux kernel version 5.1.0rc4. We run each benchmark
pinned to a single NUMA node. To avoid interference from CPU core
contention, we do not run more application threads than available
cores.

\paragraph{State-of-the-art.} We compare \sys to a number of tiered
memory management systems including Intel Optane DC 2LM~\cite{optane}
and Linux AutoNUMA~\cite{autonuma} that do not provide QoS, and
HeMem~\cite{hemem}, where we statically partition tiered memory among
applications and run a HeMem instance for each.
% \simon{Was HeMem modified to also just sample reads
  % or does it sample writes, too?} \amanda{Just reads}
% We also compare to \sys with a QoS-agnostic policy
% that evenly divides DRAM among running applications \simon{Still
%   done?}.
Both HeMem and \sys use the I/OAT DMA engine for page migration. HeMem
and \sys are configured to use 128 GB of DRAM and 768 GB of NVM as
fast and slow memory tiers, respectively, via DAX files. To generalize
our evaluation beyond Optane, we modify HeMem to use the same PEBS
counters as \sys, eliminating any Optane-specific optimizations (e.g.,
read-write asymmetry) in HeMem. 2LM and AutoNUMA use  all available
DRAM and NVM as fast and slow memory tiers, respectively. 

\subsection{Microbenchmarks}\label{sec:microbench}

We use the GUPS~\cite{gups} microbenchmark to evaluate the overheads
and performance of the various \sys policies and mechanisms. GUPS
executes parallel read-modify-write operations to eight byte memory
objects in a configurable sized memory region. Each thread does a
configurable number of operations in either a random or skewed access
pattern and measures the giga updates per second (GUPS) it
performs. % When run with
% a skewed access pattern, a variable portion of the GUPS working set
% is considered hot and accessed 90\% of the time. We allow a warmup 
% phase of 1 billion operations per thread and then measure steady-state
% throughput in GUPS. 

\subsubsection{\sys Overheads (Single Process)}\label{sec:overheads}

\setlength{\columnsep}{5pt}
\begin{figure}
  \centering
  \includegraphics[width=\columnwidth]{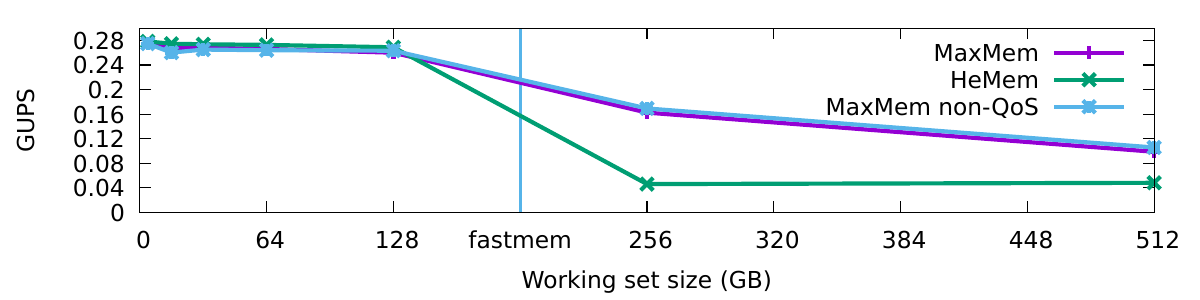}
  \caption{GUPS throughput. \simon{Nice to have: more data points
      around DRAM and one each for 320, 384, 448GB.}}
  \label{fig:eval-gups-warmset}
  % \end{wrapfigure}
\end{figure}

% \simon{Superseded by next benchmark.} \amanda{So we are skipping the
%   regular hotset benchmark completely then?} \simon{The benchmark
%   described in the next paragraph will include this benchmark. You
%   should have one configuration where the hot and warm set fit in
%   DRAM, which should be identical to this paragraph. I've rewritten
%   the text to clarify.}

% and a 512 GB working set size and vary the size of the
% hot set and report the average steady-state throughput in GUPS after a
% warmup phase in Figure \amanda{TODO}.  \amanda{results should
%   hopefully show that with a single GUPS app we get the same/almost
%   the same performance as with HeMem; similar performance graph to the
%   one in the HeMem paper with hot set GUPS}

To understand the overheads of \sys's architecture and the efficacy of
the heat gradient mechanism, we first run a single GUPS process with
16 threads and report the average steady-state throughput in
Figure~\ref{fig:eval-gups-warmset}. We configure GUPS with a hot set,
accessed 60\% of the time, a warm set, accessed 30\% of the time, and
a working set, accessed 10\% of the time, as well as a fixed size
ratio of 2$\times$ between hot and warm set and between warm and
working set. % We vary the hot set size, keeping the set size ratios
% constant. 
We configure $t_{miss} = 0.1$ to exercise \sys's QoS policy mechanisms
(\sys) and $t_{miss} = 1$ (\sys non-QoS). This parameter has no effect
in a single process scenario and serves only to break out QoS policy
overheads.
% . We also compare to a configuration of \sys that
% does not provide QoS among applications to show any overheads of
% \sys's QoS policy.

While the working set fits in DRAM, all systems attain a high average
GUPS throughput. In this case, GUPS performs at most 3\% worse with
\sys than with HeMem, indicating that the overhead for \sys's
mechanisms is at most 3\%. There is no noticeable difference between
\sys and \sys non-QoS, indicating that the QoS policy does not perturb
single process results. With a working set of 256 GB, the hot set (64
GB) and the warm set (128 GB) cannot both fit in DRAM. \sys can
prioritize the hot set for placement in DRAM due to its heat gradient
tracking. HeMem uses a hotness threshold that cannot distinguish
between hot and warm sets and suffers a lower average throughput,
about 30\% of \sys. This demonstrates the benefit of \sys's heat
gradient mechanism as the working set size expands.
 % The throughput of \sys without QoS policy is about 7\%
% better than \sys with QoS policy, demonstrating the overheads of the
% QoS policy is about 7\%.  \amanda{Removed 2LM results here since we
%   know from HeMem that they will be bad and didn't want to have too
%   many lines. Thought it would be better to show the non-QoS policy
%   here to try to tease out the overheads of QoS. Can put 2LM results
%   back if needed}

% \amanda{Any other useful single-app performance benchmarks here?}
% \simon{Potentially repeat the moving hotset benchmark from HeMem paper.}

\subsubsection{\sys QoS (Process Colocation)}\label{sec:performance}

\begin{figure*}
  \centering
  \subfloat[Instantaneous GUPS throughput.]{\label{fig:fair-gups-tput}\includegraphics[width=\textwidth]{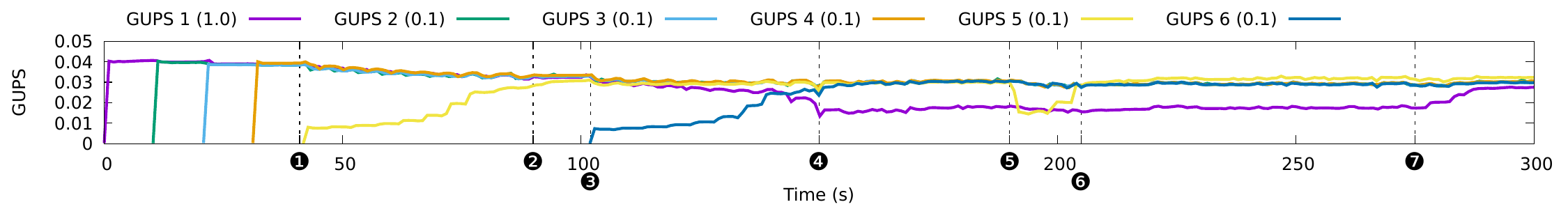}}\\
  % \qquad
  \subfloat[Instantaneous FMMRs ($a_{miss}$).]{\label{fig:fair-miss-ratios}\includegraphics[width=\textwidth]{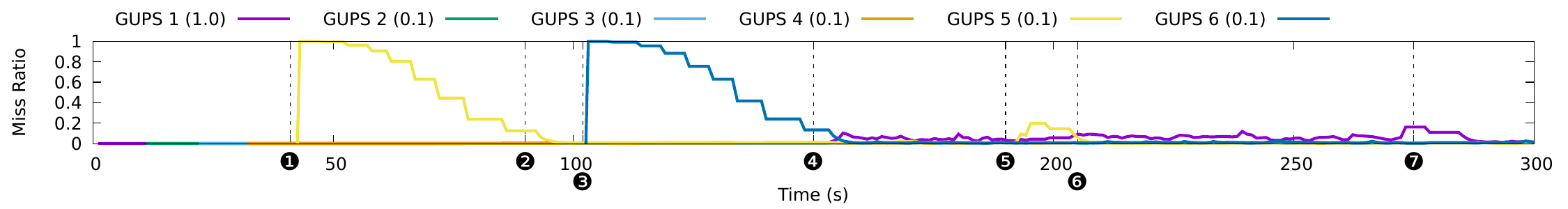}}
  \caption{Dynamic workload timeline with 6 GUPS processes.}
  \label{fig:eval-gups-colocate}
  % \end{wrapfigure}
\end{figure*}

We evaluate how well \sys provides QoS dynamically. To do so, we run
six GUPS processes. Each process runs with 2 threads and accesses a 32
GB working set. The first process is best effort, with $t_{miss} = 1$.
The next five processes each have an 16 GB hot set that is accessed
90\% of the time and they are latency sensitive, each with 
$t_{miss} = 0.1$. The first five processes all start 10 seconds apart,
while the last process starts 60 seconds after the previous process.
We plot a timeline of the per-second GUPS throughput and the 
$a_{miss}$ of each of the processes in
 Figure~\ref{fig:eval-gups-colocate}. We use this workload to
evaluate \sys's performance regarding three different dynamic QoS
scenarios: (i) dynamically arriving applications, (ii) dynamically
changing access patterns, and (iii) dynamically changing QoS
requirements.

\paragraph{Dynamically arriving applications.}
\sys dynamically reallocates DRAM among applications according to QoS
targets as applications arrive. The first four processes to start can
all fit in fast memory, so they all attain a high throughput and low
$a_{miss}$. Once the fifth process starts (\circled{1}), fast memory
is full, so this new process starts without fast memory. \sys observes
that $a_{miss} = 1 > 0.1 = t_{miss}$ for this fifth process. \sys
reallocates fast memory by taking memory from the other GUPS processes
which all have $t_{miss} > a_{miss}$. As fast memory is reallocated
from the other processes to the fifth process, the throughput of the
fifth process increases while the throughput of the others
decrease. Note that there is also a small drop in throughput of the
second, third, and fourth processes during this time, even though
these processes are not giving up any of their fast memory
allocations. Their throughput is affected by \sys migrations causing
shared memory bandwidth contention. However, the contention is limited
due to \sys's 1GB/s migration rate limit
(\S\ref{sec:policy}). $a_{miss} \leq t_{miss}$ always holds for these
processes.  At \circled{2}, the throughput and FMMRs stabilize and
\sys provides QoS to all processes, according to their target FMMRs.

At \circled{3}, the sixth process starts and the process repeats. \sys
reallocates fast memory from the other process to the sixth
process. \sys takes more memory from the first process whose
$t_{miss} = 1$ than the other processes whose $a_{miss}$ are closer
to $t_{miss}$ according to the method described in \S\ref{sec:policy}. 
As a result, the throughput of the first process degrades more than 
the throughput of the others as its fast memory allocation shrinks
compared to the rest. At \circled{4}, the system stabilizes once again 
with \sys providing QoS to all processes.

\paragraph{Dynamically changing access patterns.}
%We want to show how well \sys can meet the quality-of-service promises
%in the face of changing application access patterns. We run a
%best-effort instance of GUPS with 8 threads, a working set size of 256
%GB and a hot set size of 64 GB. We run a prioritized instance of GUPS
%with 8 threads, a working set size of 256 GB, and a hot set size of 32
%GB. After 100 seconds of running, the prioritized GUPS increases its
%hot set size to 128 GB. We plot the per-second throughput of both GUPS
%instances as well as the amount of DRAM allocated to each
%instance. Results are shown in Figure \amanda{TODO} \amanda{Here the
%  results will show good steady-state throughput of both GUPS
%  instances. Once the prioritized GUPS increases its hot set size, the
%  best-effort GUPS will start to show lowered throughput as well as
%  lowered DRAM allocations while the prioritized GUPS will show an
%  initial drop in throughput as the new elemetns of the hot set are
%  identified and migrated up to DRAM. The DRAM allocation of the
%  prioritized GUPS will increase while the DRAM allocation of the best
%  effort GUPS will decrease}
At \circled{5}, we increase the hot set size of the fifth process by
50\%. \sys's FMMR mechanism immediately detects this dynamic access
pattern change, as the FMMR of the fifth process spikes to almost
0.2. \sys now allocates more fast memory to this process. As the fifth
process is allocated more fast memory, its FMMR decreases and its
throughput increases, until all processes achieve their QoS
requirements (\circled{6}).

\paragraph{Dynamically changing QoS requirements.}
At \circled{7}, the target FMMR of the first process is updated from
1.0 to 0.1. The adjustment causes \sys to migrate more of its hot set
to fast memory as some of its hot data was accessed from NVM (seen by
its $a_{miss} > 0.1$) before this change. Fast memory is taken from
the other processes. Their cold data, according to their heat
gradients, is migrated to slow memory, minimizing the performance
impact of this reallocation. Once the system stabilizes, each process
exhibits an FMMR that is at or below their new target FMMRs, providing
QoS.

\subsection{Application Benchmarks}\label{sec:app_eval}

To demonstrate that dynamic QoS improves colocation performance, we
evaluate the performance of colocated Big Data applications with
tiered main memory management systems that embody different approaches
to QoS. In particular, we evaluate the tail latency and throughput of
FlexKVS~\cite{FlexNIC}, a high-performance key-value store, running as
a high-priority LS application alongside each of three BE
applications: GUPS~\cite{gups}, GapBS~\cite{gapbs}, and NAS
BT~\cite{bt}. We investigate both static workload configurations, as
well as a dynamically changing workload mix. To minimize CPU
interference, we pin each application to its own set of CPU
cores. FlexKVS is run with 4 threads, while GUPS and GapBS are each
run with 8 threads.

% \begin{figure*}
%   \centering
%   \subfloat[Average latency.]{\label{fig:eval- flexkvs-lat-avg}\includegraphics[width=\columnwidth]{figures/applications/background/eval_flexkvs_avg_lat.pdf}}
%   \qquad
%   \subfloat[99\% latency.]{\label{fig:flexkvs-lat-tail}\includegraphics[width=\columnwidth]{figures/applications/background/eval_flexkvs_tail_lat.pdf}}
%   % \includegraphics[width=\columnwidth]{figures/background-flexkvs/flexkvs_hemem.png}
%   \caption{FlexKVS operation latency across configurations.}
%   \label{fig:eval-flexkvs-lat}
%   % \end{wrapfigure}
% \end{figure*}

% \setlength{\columnsep}{5pt}
% \begin{figure}
%   \centering
%   \includegraphics[width=\columnwidth]{figures/applications/background/eval_flexkvs_tput.pdf}
%   \caption{FlexKVS throughput across configurations.}
%   \label{fig:eval-flexkvs-tput}
%   % \end{wrapfigure}
% \end{figure}

\paragraph{Colocation performance without tiered memory.} We first run
all applications in fast memory to establish baseline colocation
performance without tiered memory. In this scenario, the working set
size of each application is set to 64GB, such that the colocated
applications fit. We find that there is negligible interference---%
% There is a slight
% 3\% increase in average FlexKVS latency when colocated versus
% isolated, while
FlexKVS 99\%-ile latency is 34$\mu$s at 0.15 MOPS in isolation and in
colocation, and regardless of co-runner. We conclude that the baseline
system configuration without tiered memory isolates these applications
well. As configured, CPU scheduling also does not cause performance
interference.

\begin{figure}
  \centering
  \includegraphics[width=\columnwidth]{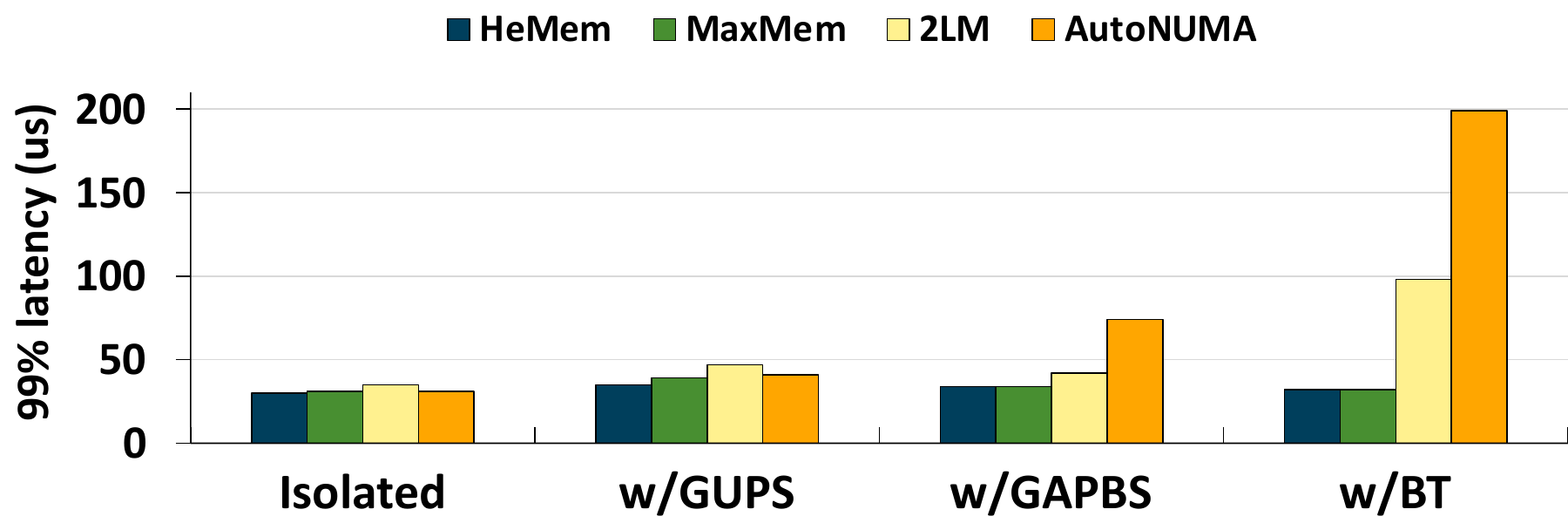}
  \caption{FlexKVS 99\%-ile latency under colocation.}%
  \label{fig:bg-flexkvs-lat}
\end{figure}

\begin{figure}
  \centering
  \includegraphics[width=\columnwidth]{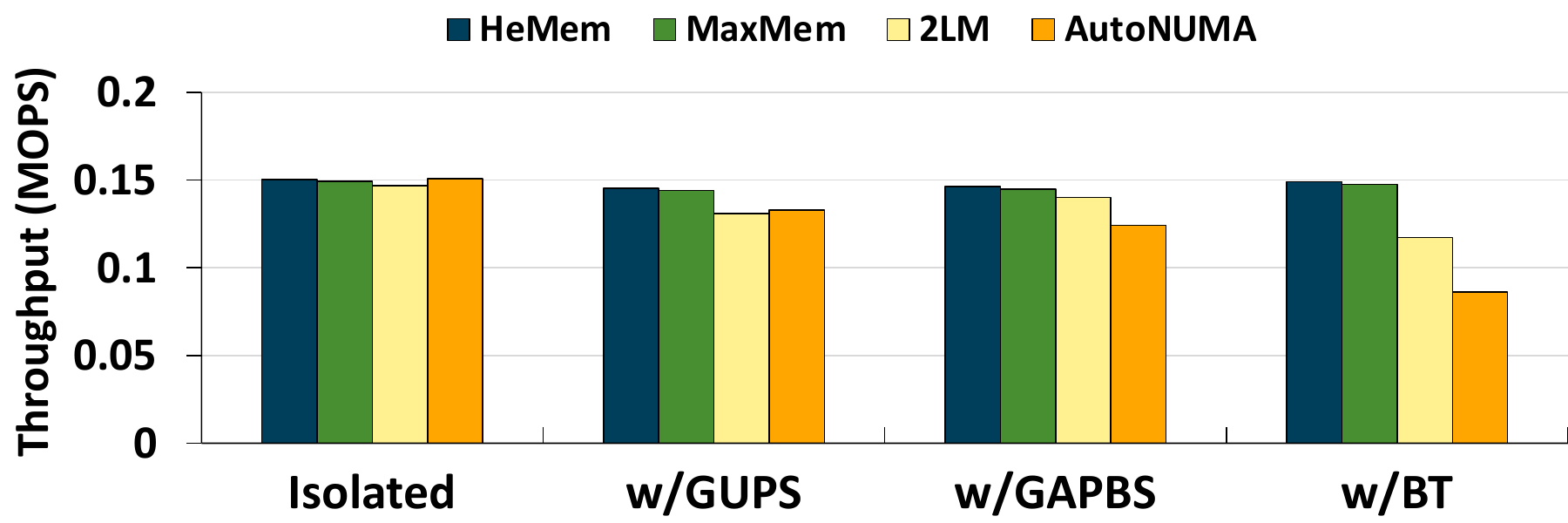}
  \caption{FlexKVS throughput under colocation.}%
  \label{fig:bg-flexkvs-tput}
  % \end{wrapfigure}
\end{figure}

\begin{figure}
  \centering
  \includegraphics[width=\columnwidth]{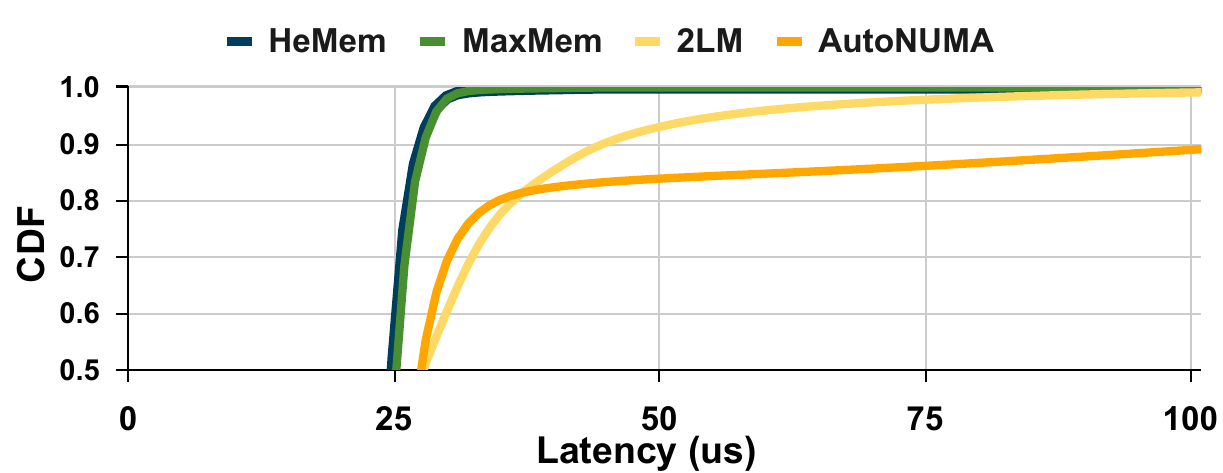}
  \caption{FlexKVS latency CDF under BT colocation.}%
  \label{fig:bg-flexkvs-bt-cdf}
\end{figure}

\paragraph{Static workload.} To measure how well each tiered memory
system meets the performance requirements of LS applications, we
compare 2LM (no QoS), AutoNUMA (no QoS), HeMem (static QoS), and \sys
(dynamic QoS). For \sys, we configure FlexKVS with $t_{miss} = 0.1$
and the BE tasks with $t_{miss} = 1$. HeMem's static QoS approach is
an upper bound for this experiment: we configure a fast memory
partition of 128GB, fitting FlexKVS's entire hot set.
%
% For HeMem, we evaluate two configurations:
% (i) A fast memory partition size of 64GB (HeMem: Underprovisioned)---80\% of
% FlexKVS's working set is in slow memory in this configuration and tail
% accesses are dominated by slow memory performance; (ii) A 128GB fast
% memory partition (HeMem: Overprovisioned)---FlexKVS's entire hot set fits
% into fast memory in this configuration. \simon{Update with new results.}
%
The results, in terms of FlexKVS tail latency and throughput, are
shown in Figures~\ref{fig:bg-flexkvs-lat} and
\ref{fig:bg-flexkvs-tput},
respectively. Figure~\ref{fig:bg-flexkvs-bt-cdf} shows a latency CDF
of FlexKVS under BT colocation. Our workloads (using Big Data
configurations \cite{hemem}) and their working set sizes in tiered
memory are described in Table~\ref{tab:apps}. We observe:

\begin{table}
  \centering
  \footnotesize
  % \small
  \begin{tabular}{llr}
    App & Workload & Workset\\
    \hline
    FlexKVS & 32B keys, 16KB vals, 9:1 read:update, 23\% hot & 320 GB\\
    GUPS & Uniform random memory update & 256 GB\\ 
    GapBS & Betweenness centrality graph algorithm & 128 GB\\
    NPB &  Block Tri-diagonal solver (BT) & 180 GB\\
  \end{tabular}
  \caption{Workloads and working sets in tiered memory.}%
  \label{tab:apps}%
\end{table}

\begin{compactenum}[1.]
\item \sys provides tail latencies and throughput comparable to
  HeMem. Figure~\ref{fig:bg-flexkvs-bt-cdf}, in particular, shows that
  HeMem and \sys tails are nearly identical for BT colocation, which
  is the most memory intensive co-runner due to its use of vector
  instructions. \sys achieves this performance by automatically and
  dynamically providing enough fast memory to FlexKVS to achieve its
  target FMMR. Note that 99th percentile latency (shown in
  Figure~\ref{fig:bg-flexkvs-lat}) is dominated by slow memory
  accesses, but still affected by bandwidth contention on slow memory
  when not enough fast memory is allocated. \sys allocates the
  necessary amount of fast memory to attain target FMMRs, causing
  minimal slow memory bandwidth contention.

\item 2LM and AutoNUMA do not perform well. For all colocations, but
  in particular for the BT colocation, 2LM and AutoNUMA exert far
  higher tail latency and lower
  throughput. Figure~\ref{fig:bg-flexkvs-bt-cdf} shows that AutoNUMA's
  and 2LM's latency deteriorates after the 80th percentile.
\end{compactenum}

%  prioritize the performance over the
% best-effort tasks, achieving competitive throughput and latency to the
% statically partitioned HeMem configurations. Like with the other
% configurations, the 99\% latency is bound by NVM accesses.

% \begin{figure}
%   \centering
%   \caption{FlexKVS 90th percentile latency.}
%   \label{fig:eval-dyn-flexkvs-lat90}
% \end{figure}

% \begin{figure}
%   \centering
%   \caption{FlexKVS 99th percentile latency.}
%   \label{fig:eval-dyn-flexkvs-lat99}
% \end{figure}

\medskip

\noindent
We conclude that, while DRAM provides good QoS among our colocated
applications in this setting and does not need special QoS support,
tiered memory systems that do not support QoS fail to isolate memory
access performance. While HeMem's static memory partitioning achieves
QoS, it is only optimal when the workload is static and the partition
is manually configured to match the hot set. \sys can achieve the
performance of HeMem dynamically and automatically finds the
appropriate fast memory allocation. In this experiment, \sys achieves
HeMem's performance while using on average 32\% less fast memory than
the manually configured HeMem partition.

% Only \sys
% supports dense colocation of LS and BE applications via dynamic QoS.

\begin{figure}
  \raggedleft
  \includegraphics[width=\columnwidth]{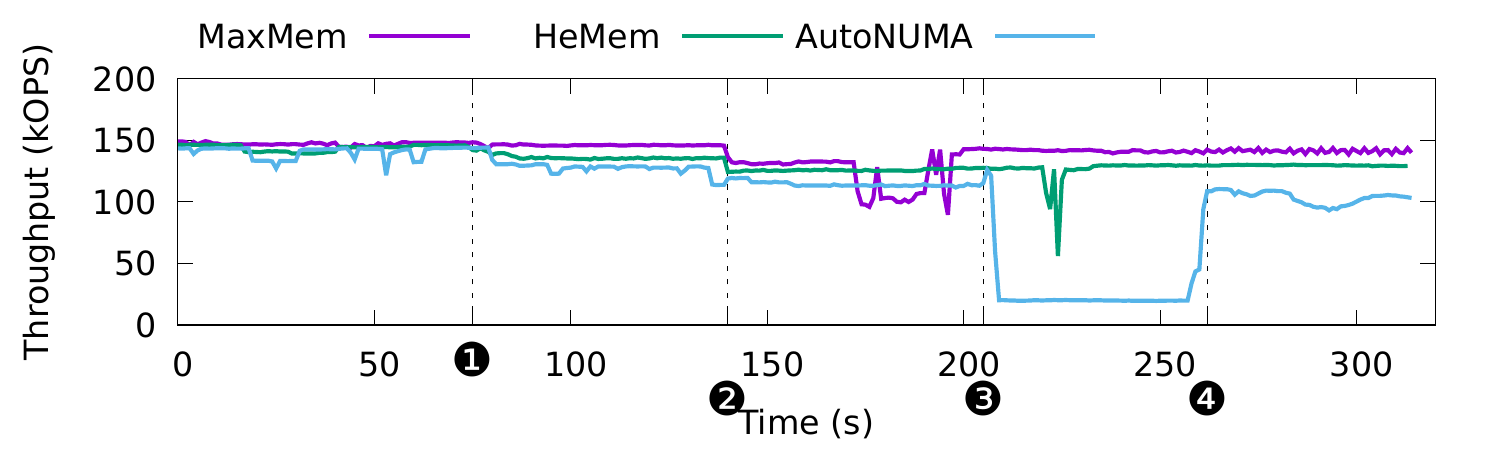}
  \includegraphics[width=.99\columnwidth]{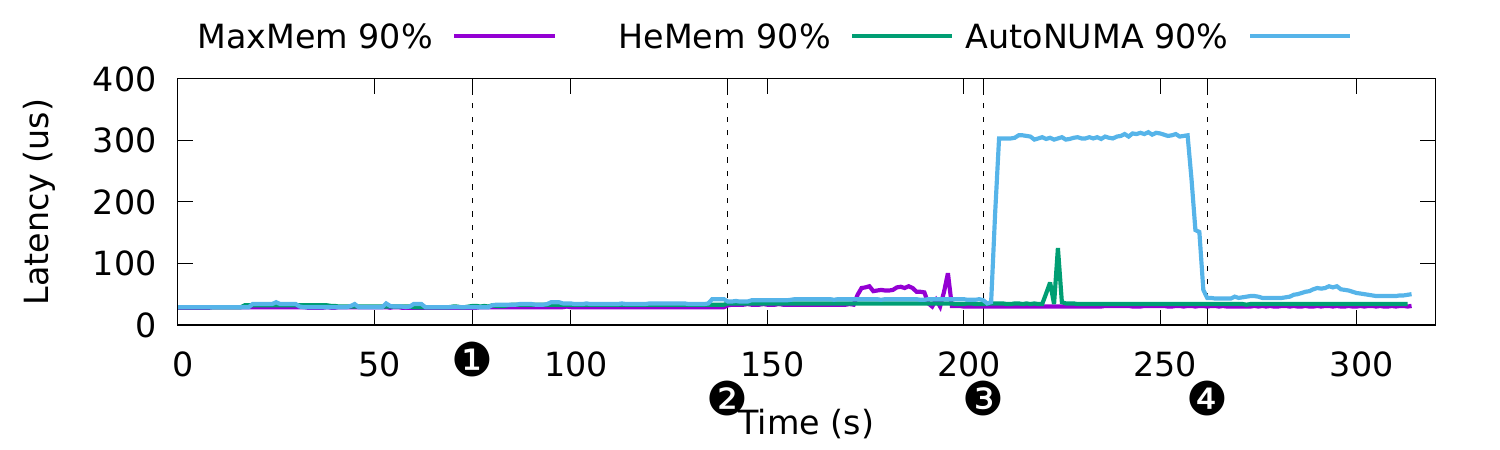}
  \includegraphics[width=\columnwidth]{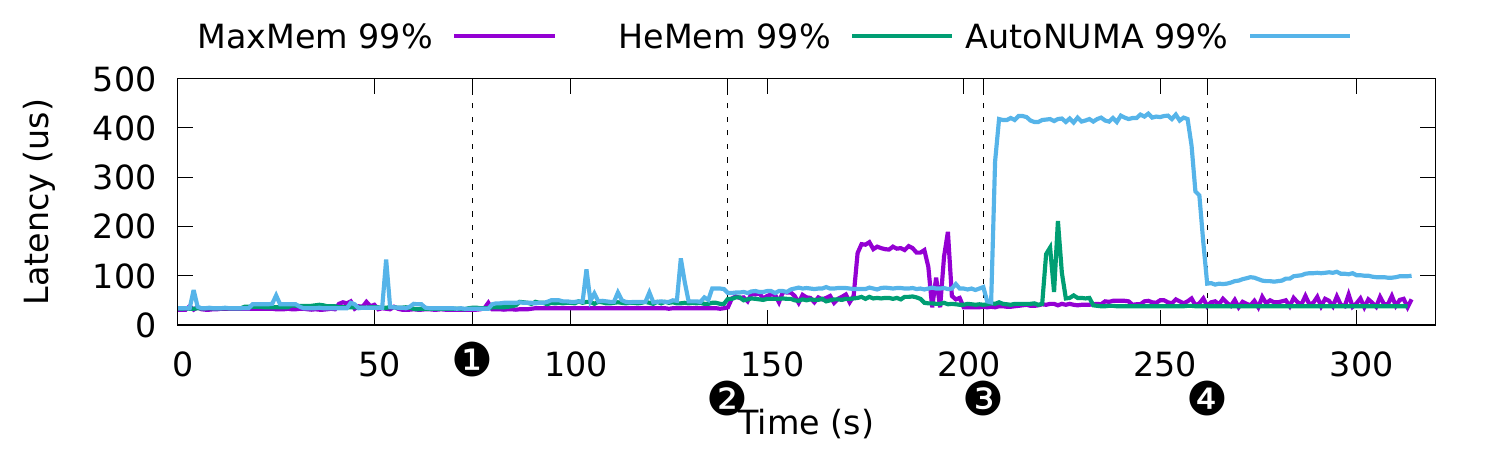}
  \includegraphics[width=\columnwidth]{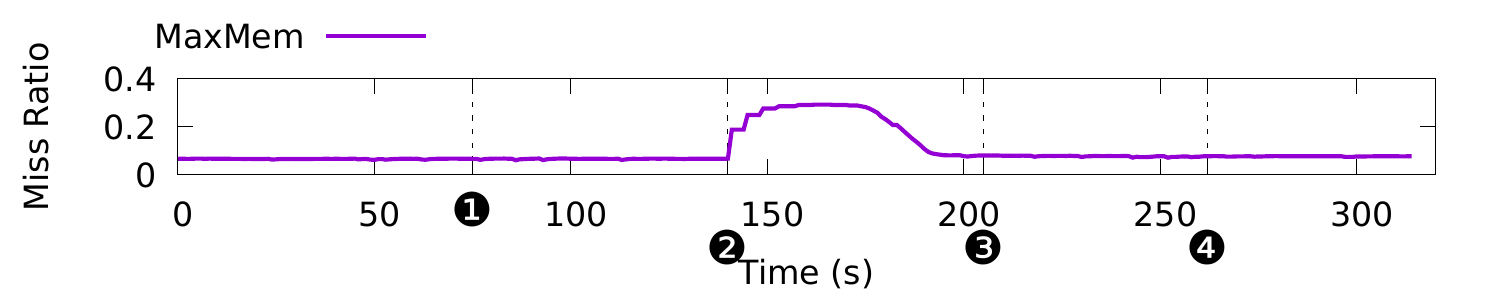}
  \caption{FlexKVS throughput, tail latencies, and $a_{miss}$.}
  \label{fig:eval-dyn-flexkvs}
\end{figure}

\paragraph{Dynamic workload.}
Next, we measure how well \sys can react to dynamically changing
workloads. We run FlexKVS as an LS application with a 320 GB working
set and a 48 GB hot set. We colocate FlexKVS with GapBS with a graph
with $2^{28}$ vertices and with uniform GUPS with a working set size
of 128 GB. For \sys, FlexKVS is given a target FMMR $t_{miss} = 0.1$
while GapBS and GUPS are given target FMMRs $t_{miss} = 1$. For HeMem,
we partition fast memory into three equal sections 42 GB in size. Both
FlexKVS and GapBS start at the same time. We allow each system a
warmup time of 100 seconds to identify a hot set and then measure the
per-second throughput, 90th, and 99th percentile latencies, as well as
$a_{miss}$ of FlexKVS in Figure~\ref{fig:eval-dyn-flexkvs}. FlexKVS
accesses its hot set 90\% of the time, so 90th percentile latencies
show how well the hot set is isolated. 99th percentile latencies show
how well the working set is isolated. Note that \sys attempts to
provide QoS only for the hot set ($t_{miss} = 0.1$). We continue
running FlexKVS and GapBS for 75 seconds before starting GUPS. We then
run for an additional 65 seconds, at which point FlexKVS increases its
hot set size to 74 GB. We then run for another 240 seconds.

For the first 75 seconds after the warmup phase, \sys exhibits 4\%
better throughput than HeMem. Tail latencies show a similar behavior.
HeMem's fast memory partition of 42 GB is not large enough to hold the
initial hot set of FlexKVS, so its throughput suffers due to the
portion of hot key accesses that are served from slow memory. Due to
the static partitioning, HeMem cannot make use of the portion of fast
memory reserved for GUPS, even though GUPS is not running. \sys
dynamically determines the fast memory partition that allows FlexKVS
to keep its hot set in fast memory and meet its target FMMRs.
AutoNUMA does not allocate enough fast memory to FlexKVS, resulting in
up to 13\% lower throughput and latency spikes of up to 3$\times$
higher 99\% latency than the other systems in this phase.

At \circled{1}, we start GUPS. HeMem and AutoNUMA exhibit a slight
drop in throughput as FlexKVS contends with GUPS over slow memory
bandwidth. FlexKVS throughput and tail latencies remain steady with
both \sys after GUPS starts. In fact, GUPS ($t_{miss} = 1$)
is allocated no fast memory with \sys when it starts up because GapBS
and FlexKVS are using the entire fast memory. As GUPS continues to 
run, AutoNUMA exhibits worse throughput and tail latency as more fast 
memory is allocated to GUPS, resulting in interference with FlexKVS.
FlexKVS with \sys exhibits 8\% and 14\% better throughput than HeMem
and AutoNUMA, respectively, and up to 27\% and 70\% lower 99\%
latencies than HeMem and AutoNUMA, respectively during this phase.

At \circled{2} we increase the hot set of FlexKVS from 42 GB to 74 GB.
HeMem's static partition of fast memory is not large enough to fit the
enlarged hot set of FlexKVS and HeMem exhibits a 8.6\% drop in FlexKVS
throughput as some of its hot accesses are now served from slow
memory. Similarly, AutoNUMA exhibits a 10\% drop in throughput at this
point.

\sys reacts to the increase in FlexKVS's FMMR and allocates more fast
memory to it.
% After an initial 9\% drop in throughput, % \sys is able to start to
% increase the fast memory allocated to FlexKVS by reallocating fast
% memory from GapBS.
Throughput and latency of FlexKVS are temporarily degraded while this
occurs. By \circled{3}, \sys has restored FlexKVS throughput. No other
system restores FlexKVS throughput and latency. AutoNUMA exhibits
substantial performance interference between \circled{3} and
\circled{4}, resulting in up to 86\% decrease in throughput compared
to \sys and up to 10$\times$ higher tail latency, respectively. HeMem
also exhibits a brief period of performance interference during this
phase, resulting in an up to 56\% drop in throughput and a 3$\times$
increase in 99\% latency. By the end of the benchmark, \sys achieves
11\% and 38\% better throughput than HeMem and AutoNUMA, respectively.

\subsection{\sys Parameter Sensitivity}\label{sec:sensitivity}

We evaluate how \sys's performance is affected by its various memory
measurement and migration parameters. We conduct this study using
FlexKVS as an LS and GapBS as a BE application. In particular, we run
FlexKVS for 30 seconds with a hot set that fits in fast memory. Then,
we double the size of FlexKVS's hot set and observe how quickly \sys
restores FlexKVS's FMMR, as well as how FlexKVS's operation tail
latency is affected by \sys migration.

\begin{figure}
  \centering
  \includegraphics[width=\columnwidth]{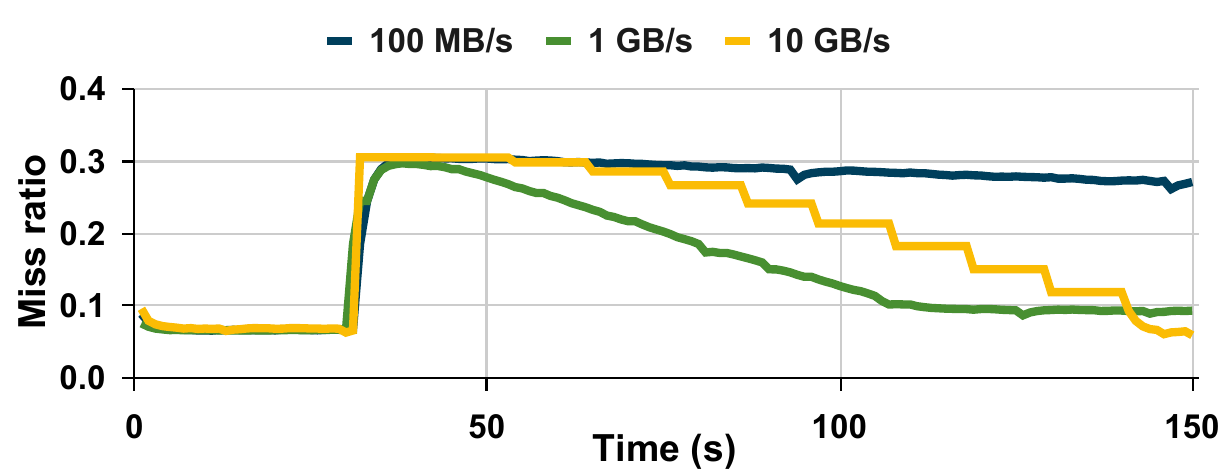}
  \caption{FlexKVS FMMR varying migration rate.}%
  \label{fig:sens-missrate}
\end{figure}

\begin{figure}
  \centering
  \includegraphics[width=\columnwidth]{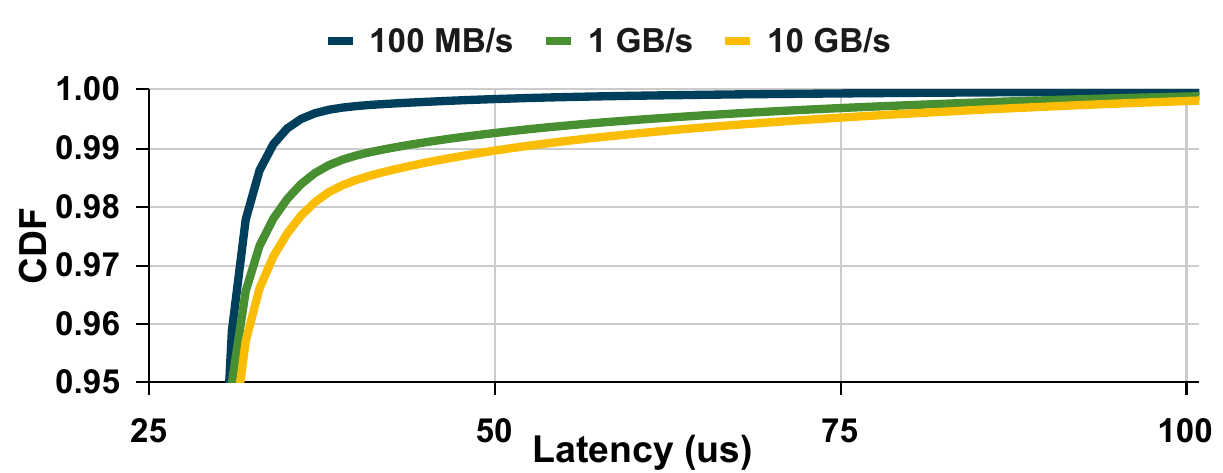}
  \caption{FlexKVS latency CDF varying migration rate.}%
  \label{fig:sens-lat-cdf}
\end{figure}

\paragraph{Memory migration rate.} We analyze a variety of memory
migration rate caps. Figure~\ref{fig:sens-missrate} shows that 1 GB/s
restores FlexKVS's FMMR the quickest. 100 MB/s is migrating memory too
slowly, while 10 GB/s is requesting a higher migration rate than
available migration bandwidth. This stalls \sys's policy thread during
migration, as is evident from the resulting step function, causing
\sys to take longer to converge. Figure~\ref{fig:sens-lat-cdf} shows a
CDF of FlexKVS operation latency. We can see that migration affects
95th percentile and higher latencies. Higher migration rates have a
larger effect, as expected. Nevertheless, \sys's migration rate used
in this paper is 10 GB/s, in line with HeMem's~\cite{hemem}.

\paragraph{Migration epoch duration.} We analyze a range of migration
epoch durations---100ms, 500ms, 1s, and 2s, while fixing migration
rate at 1 GB/s. We find that 1s provides the best FMMR convergence
time and reaction time. FlexKVS tail latencies are negligibly affected
by epoch duration. Short migration epochs migrate too few pages at a
time, limiting migration performance. Long migration epochs increase
reaction times, prolonging time to convergence.

\section{Related Work}\label{sec:related}

% We describe here further related work, beyond \S\ref{sec:tieredmem}.

\paragraph{Tiered memory systems} Further tiered memory
systems~\cite{heteroos,nimble,thermostat,hemem,autotiering-NVM,xmem,unimem,2pp,clio,fastswap,aifm}
do not maximize server utilization or provide QoS for Big Data
applications. These systems propose a variety of techniques to
determine which tier to use for application data. To be lightweight,
\sys builds on HeMem~\cite{hemem} and uses PEBS for this task.

%  both for NVM and disaggregated memory
% \simon{Presumably, none of these provide QoS. Edit this down to just
%   say that, rather than go into the details.}.
% HeteroOS~\cite{heteroos}, Nimble~\cite{nimble},
% Thermostat~\cite{thermostat}, HeMem~\cite{hemem},
% AutoTiering~\cite{autotiering-NVM}, and others~\cite{xmem, unimem,
%   2pp} consider NVM as the slow memory tier while AIFM~\cite{aifm},
% Clio~\cite{clio}, and FastSwap~\cite{fastswap} consider remote,
% disaggregated memory as the slow memory tier. These systems use a
% variety of techniques, such as page faults and page table access bit
% scanning~\cite{heteroos, nimble, thermostat, autotiering-NVM,
%   fastswap, software-far-mem}, specialized memory APIs~\cite{xmem,
%   unimem, 2pp, clio, aifm}, or hardware performance
% counters~\cite{hemem, pond, tpp} to determine which tier to place a
% given block of memory. \sys builds on HeMem~\cite{hemem} and uses
% hardware performance counters to determine which tier to place a page
% of memory in.

\paragraph{QoS in other contexts} QoS is a full-stack concern and is
thus addressed across the hardware/software stack. For example,
distributed caching engines for web applications provide QoS among
tenants via partitioning~\cite{cachelib}, with coarse-grained input
via SLAs and cache utilization telemetry. As with static tiered memory
partitioning in HeMem~\cite{hemem}, these techniques often result in
underutilization of cache partitions due to bursty
workloads. Quantitative methods, including cache miss ratio, have been
studied but usually discarded due to space overheads. OSCA~\cite{osca}
is a recent block storage cache that tackles space overheads by using
re-access ratio to obtain the cache requirements, using total hit
traffic as the optimization target, and searching for an optimal cache
partition using dynamic programming. Memshare~\cite{memshare} is a
multi-tenant KV cache that dynamically repartitions memory among
tenants via a segmented in-memory log. Pisces~\cite{pisces} provides
performance isolation and max-min fairness in multi-tenant cloud
storage via smart placement and weighting of storage partitions. \sys
has to be lighter-weight than these approaches, while supporting
multi-processing. \sys thus samples FMMRs via a trusted central
manager process.

Intel introduced cache allocation technology (CAT)~\cite{cat} into
recent CPUs to provide CPU cache QoS. CAT allows system administrators
to group processes and VMs into service classes that are allocated a
partition of last-level cache (LLC). Partitions may be altered at run
time. However, the OS is responsible for implementing appropriate
partitioning policies and measurements. \sys provides such policies
and measurements and it is conceivable that the LLC may be integrated
as an additional memory tier.

Virtual machines implement memory capacity sharing via ballooning and
employ share-based memory allocation to VMs \cite{esx}. These
mechanisms consider only a single tier of memory and are thus focused
on the working set of VMs, rather than memory heat gradients.

\paragraph{User-space OS services} \sys's user-space design takes
inspiration from recent user-space OS service proposals, including
ghOSt~\cite{ghost}, a user-space Linux CPU scheduler, Snap~\cite{snap}
and TAS~\cite{tas}, user-space networking stacks, and
Strata~\cite{strata} and Assise~\cite{assise}, user-space
(distributed) file systems. User-space development is fast and
flexible, unlike error-prone kernel-space development, but user-space
performance overheads may be larger and specific techniques have to be
adopted to stay efficient. Each system developed its own techniques,
specific to each task, to do so. \sys leverages a split library and
central manager design, as well as batched samples in shared memory
buffers via PEBS, asynchronous relay of page faults to user-space via
userfaultfd, as well as batched migration via DMA offload, to make
user-space management of tiered memory efficient.

\section{Conclusion}

We present \sys, a tiered main memory management system that maximizes
Big Data application colocation and performance. \sys uses an
application-agnostic and lightweight memory occupancy control
mechanism based on FMMRs to provide application QoS under increasing
colocation. By relying on memory access sampling and binning to
quickly identify per-process memory heat gradients, \sys maximizes
performance for many applications sharing tiered main memory
simultaneously. \sys is designed as a user-space memory manager to be
easily modifiable and extensible, without complex kernel code
development. Our evaluation confirms that \sys provides up to 76\% and
80\% lower 90th and 99th percentile tail latency, while providing 11\%
better throughput for a Big Data key-value store in consolidated
workload mixes than the next best solution.

{
%\footnotesize
%\setlength{\bibsep}{1pt}
\bibliographystyle{ACM-Reference-Format}
\interlinepenalty=10000
\bibliography{bibliography}
}

\end{document}